\documentclass[%
aip,
amsmath,amssymb,
reprint,%
]{revtex4-1}

\usepackage{graphicx}
\usepackage{dcolumn}
\usepackage{bm}

\usepackage[utf8]{inputenc}
\usepackage[T1]{fontenc}
\usepackage{mathptmx}
\usepackage{etoolbox}

\makeatletter
\def\@email#1#2{%
	\endgroup
	\patchcmd{\titleblock@produce}
	{\frontmatter@RRAPformat}
	{\frontmatter@RRAPformat{\produce@RRAP{*#1\href{mailto:#2}{#2}}}\frontmatter@RRAPformat}
	{}{}
}%
\makeatother

\usepackage{amsmath, amsthm, amssymb, amsfonts}
\usepackage{subcaption}
\usepackage[utf8]{inputenc}	
\usepackage[T1]{fontenc}	
\usepackage{xcolor}		
\usepackage[colorlinks = true,
linkcolor = purple,
urlcolor  = blue,
citecolor = cyan,
anchorcolor = black]{hyperref}	
\usepackage{booktabs} 		
\usepackage{nicefrac}		
\usepackage{microtype}		
\usepackage{lineno}		
\usepackage{float}			

\usepackage{lipsum}		

\usepackage{newfloat}
\DeclareFloatingEnvironment[name={Supplementary Figure}]{suppfigure}
\usepackage{sidecap}
\sidecaptionvpos{figure}{c}

\usepackage{graphicx} 
\usepackage{dcolumn} 
\usepackage{bm} 

\usepackage{mathptmx}
\usepackage{etoolbox}
\usepackage{enumerate}
\usepackage{caption}
\usepackage{comment} 
\usepackage{braket}
\usepackage{float}

\usepackage{mathdots}
\usepackage[normalem]{ulem}

\usepackage{stackengine}

\usepackage{tikz}

\definecolor{lime}{HTML}{A6CE39}
\DeclareRobustCommand{\orcidicon}{
	\begin{tikzpicture}
		\draw[lime, fill=lime] (0,0)
		circle [radius=0.16]
		node[white] {{\fontfamily{qag}\selectfont \tiny ID}};
		\draw[white, fill=white] (-0.0625,0.095)
		circle [radius=0.007];
	\end{tikzpicture}
	\hspace{-2mm}
}
\foreach \x in {A, ..., Z}{\expandafter\xdef\csname orcid\x\endcsname{\noexpand\href{https://orcid.org/\csname orcidauthor\x\endcsname}
		{\noexpand\orcidicon}}
}

\begin{document}
	
	\preprint{AIP/123-QED}
	
	\title{Analytical Performance Evaluation of Quantum Radar Architectures: From Single-Photon to Entangled-Noise Radars}
	\author{H. Allahverdi}
	\affiliation{ 
		Quantum Remote Sensing Lab, Quantum Metrology Group, Iranian Center for Quantum Technologies (ICQT), Tehran, Tehran 15998-14713, Iran
	}
	\affiliation{ 
		Laser and Plasma Research Institute, Shahid-Beheshti University, Tehran, Tehran 19839-69411, Iran
	}
	
	\author{Ali Motazedifard}%
	\homepage{https://profile.ut.ac.ir/~alimotazedifard/network}
	\email{alimotazedifard@ut.ac.ir. }
	\affiliation{ Department of Physics, 
		 University of Tehran, Tehran  14395‑547, Iran
	}%
	\affiliation{
		Quantum Remote Sensing Lab, Quantum Metrology Group, Iranian Center for Quantum Technologies (ICQT), Tehran, Tehran 15998-14713, Iran
	}%

	\date{\today}
	
	\begin{abstract}
		This article presents a comprehensive analysis of {\color{black}two classes of} quantum radars, {\color{black}including} quantum direct-detection and quantum-entangled noise {\color{black}radars}. 
		In the first case, inspired by the well-established concept of single-photon LiDARs, we {\color{black}investigated} {\color{black}the performance of} single-photon {\color{black}radars, in which} state-of-the-art single microwave-photon detectors {\color{black}are employed} to enhance the detection sensitivity and enable the detection of weaker signals. 
		We {\color{black}derived} analytical expressions for the maximum detection range {\color{black}of both classes of quantum radars} in terms of {\color{black}the} Lambert W function, {\color{black}by considering} all relevant system, target, and environmental parameters.
		{\color{black}Our} formulation facilitates direct comparison {\color{black}of noise radars} with direct-detection {\color{black}radars}, and suggests that {\color{black}a} quantum-entangled noise radar can be regarded as an enhanced direct-detection radar with an \textit{effective threshold signal-to-noise ratio}.
		Furthermore, we {\color{black}applied} this framework to classical-correlated noise radars and {\color{black}defined} the {\color{black}parameter} \textit{range enhancement factor} {\color{black}(REF)} to quantify the superiority of  quantum-entangled {\color{black}noise radars} over {\color{black}their} classical counterparts. 
		{\color{black}Moreover, we} {\color{black}introduced} a rule-of-thumb for approximating the {\color{black}REF}.
		We also {\color{black}examined} the influence of limitations imposed by various microwave detection technologies. 
		Our analysis shows that {\color{black}the} conventional antennas limit the potential benefits of quantum-{\color{black}entangled noise} radar systems. {\color{black}We also demonstrated that} the optimal detection method for these radars is {\color{black}a} microwave detector based on a quantum transducer combined with a single optical-photon detector. 
		{\color{black}We showed that, with the current technology,} implementing a quantum-entangled noise radar with {\color{black}the maximum detection range} in the order of few kilometers is possible.
		Finally, we explored the potential applications of quantum-entangled {\color{black}noise} radars.
	\end{abstract}
	
	\maketitle

\section{\label{sec:intro} Introduction}
The detection and characterization of nearby objects using microwave electromagnetic (EM) fields—commonly referred to as RADAR (Radio Detection and Ranging)—originated from the pioneering work of the German physicist Heinrich Hertz~\cite{historyRadar}. The operating principle of radar is conceptually straightforward: a microwave signal is transmitted toward a region where a target may be present, and the returned signal is analyzed to determine the presence of the target and estimate its characteristics, such as an image, range, azimuth, and velocity. Nowadays, radar systems are widely employed across various domains, including biomedical sensing~\cite{radarBook2,wirelessBioRadar}, space and CubeSat technologies~\cite{cubeSat}, navigation~\cite{radarnavigationBook}, aviation safety~\cite{radarNetworks}, and smart wearable devices~\cite{smart}.

The traditional radar systems—often referred to as direct-detection radars—rely on the analysis of the received signal power, $P_r$, as the detection function~\cite{skolnikBook}. However, their performance can be degraded significantly in the presence of unwanted interference, such as environmental thermal noise and external microwave signals at similar frequencies.
To mitigate these challenges, engineers {\color{black}have} introduced an alternative strategy: store a reference copy of the transmitted signal (known as the idler) and compare it with the received signal~\cite{narayananNoiseRadarDesign,lukinNoiseRadarTechnology,noiseRadarOverview,efficientProcessingNoiseRadar2018}. These systems, {\color{black}which are well-known as} noise radars, utilize the correlation between the received signal and the retained idler as the detection function. {\color{black}{\color{black}A} stronger correlation {\color{black}yields} more {\color{black}accurate} target detection or {\color{black}a} larger practical radar range {\color{black}under} the same conditions and system parameters}.
This leads to a fundamental question: \textit{What is the ultimate limit of signal-idler correlation in noise radar systems}?
From {\color{black}the} classical standpoint, the maximum correlation occurs when the transmitted signal and the retained idler share identical waveforms~\cite{narayananNoiseRadarDesign}, differing only in their quantum fluctuations which {\color{black}remain} distinct.
However, {\color{black}a} quantum {\color{black}mechanical solution to achieve correlations that exceed classical bounds is {\color{black}exploiting} a pair of entangled electromagnetic fields as {\color{black}the} signal-idler modes} \cite{torrome2023advances,Shapiro2020story,Pirandola2018advance}. {\color{black}This is at the heart of} {\color{black}the} quantum illumination (QI) protocol \cite{quantumilluminationSethLloyd2008}. Note that in the case of entangled {\color{black}state} signals, not only their mean fields {\color{black}(in the X- or P-quadrature) are quantum mechanically} correlated, but their {\color{black}zero-point} quantum fluctuations are also inherently correlated. {\color{black}This leads to stronger {\color{black}correlations than those achievable by} any classically correlated signals.}

The first experimental demonstration of QI was achieved in the optical domain using bulk nonlinear optics \cite{lopaevaQuantumIllumination2013}. {\color{black}In this experiment}, a $\rm 6~dB$ improvement in the error-probability exponent over the best classical strategy was demonstrated. The extension of QI to the microwave domain was proposed {\color{black}recently} \cite{barzanjehQIlluminationOMS2015}, where an electro-opto-mechanical system was proposed for generating entangled microwave–optical photon pairs.
Despite its advantages, the original QI protocol requires a \textit{joint} measurement of the returned signal and the retained idler. {\color{black}This} approach becomes impractical in radar systems where the target range (and thus the signal's arrival time) is unknown. For this reason, the system proposed in {\color{black}Ref.}~\cite{barzanjehQIlluminationOMS2015} is not directly suitable for radar applications.
The experimental demonstration of the first microwave QI-inspired radar that does not require joint measurement was shown in {\color{black}Refs.}~\cite{wilsonMicrowaveQuantumRadar2018,snakeBalaji2018}. {\color{black}The implemented system}, {\color{black}which is} known as the quantum two-mode squeezed (QTMS) radar, utilizes a Josephson Parametric Amplifier (JPA) cooled {\color{black}to below} $\rm 10~mK$ {\color{black}in} a cryostat to generate entangled microwave signals at GHz frequencies.

In recent years, significant theoretical and experimental {\color{black}progress has} been made in the field of quantum illumination (QI), both in the optical domain~\cite{helmy40dBQuLiDARNature2022,helmy100dBQuLiDARNature2023,quantumDopplerLidar2022,jeffers1QuantumRangingJamming,jeffers2QuantumRangingJamming,quantumLidarFMCW,quantumSecuredLiDAR,twoPhotonLiDAR2022,jeffers3MimicQillumination,karsa2020quantum} and in microwave QTMS radar systems~\cite{barzanjehExperimentQuantumRadar2020,balajiCorrelationExperiment,vitaliQradarExperiment2023,vitaliQradarExperiment2022TechRxiv}. 
Most of the existing {\color{black}works} on microwave QI {\color{black}have} focused on {\color{black}the} enhancement of signal-idler correlations {\color{black}via} entanglement {\color{black}{\color{black}and by improving} system performance, for example, by considering} receiver operating characteristic (ROC) curves, detection signal-to-noise ratio (SNR), {\color{black}and} measurement time. However, despite their importance, these metrics are not sufficient to evaluate the practical performance of such {\color{black}radar} systems in long-range detection scenarios. A more comprehensive analysis rooted in range-dependent system behavior is required.

{\color{black}A} standard method to evaluate {\color{black}the maximum range at which a QTMS radar can detect a target} is through a range equation that expresses the signal-idler correlation as a function of target {\color{black}range, $R$}. In Ref.~\cite{balaji2022PerformancePrediction}, an explicit expression for the Pearson correlation coefficient $\rho$ as a function of range $R$ for noise radar systems was derived.
The formulation introduced in Ref.~\cite{balaji2022PerformancePrediction}, despite its novelty, overestimated the Pearson correlation coefficient of the generated entangled signal-idler fields. It is {\color{black}valid} in the special case where an RF amplifier used to amplify the signal and idler {\color{black}modes} adds {\color{black}substantial} noise. {\color{black}This condition is not {\color{black}satisfied in QI experiments,} since {\color{black}they usually employ} low-noise amplifiers}.
{\color{black}Moreover}, their analysis neglects key {\color{black}real-world parameters, such as} atmospheric absorption and detection time. {\color{black}The latter is important for moving objects, while absorption becomes practically important in long-range applications.}
These omissions compromise the validity of the range estimation based on the formulation proposed in \cite{balaji2022PerformancePrediction}, {\color{black}leading to} an overestimation of the system's operational performance.

Some recent studies~\cite{rangeEquation2023,bischeltsrieder2024engineering,livreri2024josephson, bistaticNoiseRadar2023} have attempted to estimate the maximum detection range of quantum{\color{black}-entangled noise} radars using the classical radar range equation that was developed for direct-detection radar systems. These models assume {\color{black}power of the received signal} as the detection function.
{\color{black}This assumption is} fundamentally inconsistent with the operational principle of quantum radars, where the signal-idler correlation—not the signal power—acts as the primary detection function. Additionally, these works {\color{black}neglect} atmospheric attenuation, which can significantly {\color{black}affect the validity of their results} in  {\color{black}long-range real-world} applications.

Motivated by the above-mentioned investigations, in this paper, we aim to address these gaps by developing an analytical framework to evaluate the {\color{black}maximum detection range of quantum-entangled noise} radars.
Specifically, we {\color{black}derived} a closed-form expression for the maximum detection range of continuous-wave (CW) quantum-entangled noise radars {\color{black}in terms of the Lambert W function. 
Our analysis} takes into account the effects of atmospheric absorption, detection time, and {\color{black}signal} bandwidth. {\color{black}We also investigate how different microwave detection technologies affect the performance of quantum-entangled noise radars.} 
Our approach extends and generalizes the analytical formulation presented in Ref.~\cite{balaji2022PerformancePrediction} to include these critical practical considerations.
{\color{black}By utilizing this formulation, we {\color{black}showed} that implementing a quantum-entangled radar system with {\color{black}{\color{black}a} maximum detection range {\color{black}on the} order of kilometers} is achievable based on current technologies. 
Furthermore, we {\color{black}addressed} strategies for overcoming {\color{black}the} present limitations, such as {\color{black}increasing the} entanglement strength, {\color{black}using} single microwave-photon detectors (SMPDs), and most importantly, {\color{black}increasing} the bandwidth of the generated entangled {\color{black}microwave fields}.}
The latter is shown to significantly improve {\color{black}the} quantum advantage by reducing the photon number per mode, thus enhancing the feasibility of long-range operation {\color{black}which is} consistent with the recent experimental insights~\cite{vitaliQradarExperiment2023,vitaliQradarExperiment2022TechRxiv}. 
{\color{black}In addition, building upon} recent advances in high-efficiency SMPDs, we proposed the concept of the single-photon radar (SPR), {\color{black}which is} {\color{black}analogous to single-photon LiDARs}.

The paper is organized as follows. In Sec.~\ref{sec2:direct_detection_radars}, we review the range equation in conventional classical direct-detection {\color{black}radars}. {\color{black}Here,} we derived an analytical expression for maximum detection range and introduce the single-photon direct-detection radar in this context. We then introduce noise radars in Sec.~\ref{sec3.NoiseRdar}. Utilizing {\color{black}results} obtained in Sec.~\ref{sec3.NoiseRdar}, we compare {\color{black}the} {\color{black}performance} of QTMS radars and classical noise radars in this section. Experimental discussion on the feasibility and comparison as well as addressing challenges {\color{black}are} provided in this section. Finally, concluding remarks and outlooks for future works are discussed in Sec.~\ref{sec6.Results and Discussion}.


\section{Quantum Direct-Detection Radars}
\label{sec2:direct_detection_radars}
In this section, we first review the operational principles of direct-detection radars to {\color{black}provide an explicit expression for the maximum detection range} of SPRs. {\color{black}This} establishes a baseline for comparison with quantum{\color{black}-entangled noise} radars.

\subsection{A Short Review on Principles of Direct Detection Radar Systems} \label{subsec2-1_review}
\begin{figure}
	\centering
	\includegraphics[width=200pt]{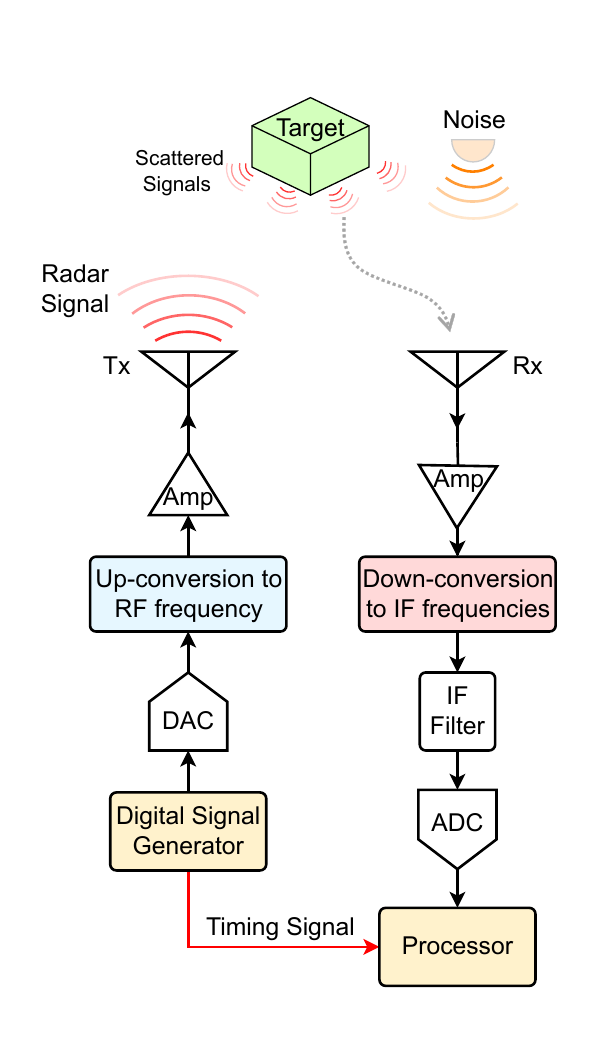}
	\caption{\label{fig:epsart}{\color{black}The} schematic diagram of a mono-static direct-detection radar system. Amp.: Amplifier; DAC: Digital-to-analog converter; ADC: Analog-to-digital converter; {\color{black}Tx/Rx: {\color{black}Transmitter/receiver} microwave antenna.}}
	\label{Fig1}
\end{figure}

Figure~\ref{Fig1} illustrates the schematic {\color{black}diagram} of a direct-detection radar system. The system begins with a digital signal generator producing an intermediate frequency (IF) signal. It is subsequently converted into an analog signal by a digital-to-analog converter (DAC). This {\color{black}analog} signal is then up-converted to radio frequency (RF), amplified, and finally transmitted toward the target region via the radar {\color{black}transmitter} antenna {\color{black}(Tx)}.
The transmitted microwave signal, with power $P_t$, undergoes atmospheric attenuation before interacting with the target. A portion of this signal is back{\color{black}-scattered} toward the radar {\color{black}receiver} antenna {\color{black}(Rx)}. At the receiver, the back{\color{black}-scattered} signal (with power $P_r$) along with {\color{black}the} environmental noise is {\color{black}first} amplified {\color{black}and} then down-converted to an {\color{black}intermediate frequency (IF)}.
The signal then passes through an IF bandpass filter with bandwidth $B_{\mathrm{IF}}$ before being digitized by a high-speed analog-to-digital converter (ADC). The digitized signal is finally processed and interpreted by a digital processor (e.g., an FPGA).
The relationship between the transmitted power $P_t$ and {\color{black}the} received power $P_r$ is governed by the classical radar range equation{\color{black}, which is given by} \cite{skolnikBook}
\begin{eqnarray}
	P_r=\eta (R) P_t , \label{power1}
\end{eqnarray}
in which $\eta(R)$ is the transfer function of the transmitter-target-receiver channel and is given by \cite{skolnikBook}
\begin{eqnarray}
	\eta(R)= \frac{\sigma G A_e}{(4\pi)^2 R^4} F(R)^2. \label{eta1}
\end{eqnarray}
{\color{black}In this equation,} $R$ is the range to the target, $\sigma$ is the radar cross-section (RCS), and $G$ {\color{black}represents} the antenna gain. {\color{black}Moreover,} $A_e$ is the effective aperture of the receiver antenna and relates to the physical antenna aperture $A$ via $A_e=\epsilon_a A$, with $\epsilon_a$ being the antenna aperture efficiency. {\color{black}Finally,} $F(R)=10^{-\gamma R/10}$ denotes the form factor describing the attenuation of the radar signal due to atmospheric absorption or loss, {\color{black}where} $\gamma$ {\color{black}is} the atmospheric absorption coefficient in units of $\mathrm{dB/m}$.

In addition to the {\color{black}portion of the radar} signal back{\color{black}-scattered} from the target, the radar receiver also collects unwanted signals from {\color{black}the} environment, {\color{black}whose power we denote} by $P_n$. When the target detection process is carried out over $M$ independent measurements, the effective signal-to-noise ratio (SNR) for the direct-detection radar system is given by $\mathrm{SNR}_{\mathrm{eff}}^{\mathrm{DDR}} = M P_r/P_n = M \eta(R) P_t/ P_n$ \cite{skolnikBook}.
It is important to note that the number of measurements $M$ depends on the integration time $\tau_{\mathrm{int}}$ and the detection bandwidth $B_{\mathrm{det}}$, through the relation $M = \tau_{\mathrm{int}} B_{\mathrm{det}}$. The detection bandwidth itself is determined by the IF filter bandwidth $B_{\mathrm{IF}}$ at the receiver, i.e., $B_{\mathrm{det}}=B_{\mathrm{IF}}$.

A direct-detection radar {\color{black}enables {\color{black}reliable identification}} of the target when the effective signal-to-noise ratio exceeds a predefined detection threshold, denoted as $\mathrm{SNR}_{\mathrm{th}}$. {\color{black}In other words}, {\color{black}reliable} detection occurs if {\color{black}and} only if $\mathrm{SNR}_{\mathrm{eff}}^{\mathrm{DDR}} > \mathrm{SNR}_{\mathrm{th}}$, {\color{black}which} {\color{black}is} commonly referred to as the detection criterion \cite{skolnikBook}.
{\color{black}Note that the threshold SNR} refers to the minimum required power of the returned radar signal, $P_{r,\mathrm{min}}$, or, {\color{black}equivalently, the} minimum detectable signal (MDS), for which the radar system {\color{black}is capable of recognizing} {\color{black}the presence of the target}. {\color{black}This leads to} ${\mathrm{SNR}_{\mathrm{th}}}=P_{r,\mathrm{min}}/P_n$.
For instance, for a typical radar system with $\mathrm{SNR}_{\mathrm{th}}=\rm 10~dB$, the power of the returned signal $P_r$ should be at least $10~\mathrm{dB}$ higher than environmental noise power $P_n$ {\color{black}so that the} radar system is able to identify the target, i.e., $P_r~(\mathrm{dB})\ge P_n~(\mathrm{dB}) + 10~\mathrm{dB}$, or equivalently, $P_r~(\mathrm{SI})\ge 10\times P_n~(\mathrm{SI})$.

By substituting the direct detection radar range equation given {\color{black}by} Eq.~\ref{power1} in the detection criterion, 
and considering that the equality in the detection criterion holds for the maximum detection range of the radar system $R_{\rm max}${\color{black}, one can get to}
\begin{eqnarray} \label{max_range_equation}
	R_{\mathrm{max}}  \mathrm{exp}\bigg(\frac{\rm ln(10)}{20} \gamma R_{\mathrm{max}}\bigg) = \bigg(\frac{M}{\mathrm{{SNR}_{th}}}\bigg)^{1/4}  R_c^{\rm (DDR)}.
\end{eqnarray}
{\color{black}Here,} $ R_c $ is the characteristic-range of the direct detection radar system and defined as
\begin{eqnarray} \label{r_c_direct} 
	R_c^{\rm (DDR)} \equiv \left( \frac{\sigma G A_e P_t}{(4\pi)^2 P_n} \right)^{1/4}.  
\end{eqnarray}
{\color{black}$R_c^{\rm DDR}$ is the} range for which the total power at the receiver for in-vacuo signal propagation ($\gamma=0 $) is $\rm 3~dB$ above the noise level, or equivalently, twice the noise power $ P_n $. 
Alternatively, it can be interpreted as {\color{black}the} range for which the SNR  at the receiver equals 1 (or 0 dB ). 
The characteristic range depends on the  {\color{black}radar} system parameters, {\color{black}the} transmitted signal power $P_t$, and the environmental thermal noise power $P_n$ {\color{black}which is} given by \cite{skolnikBook}
\begin{eqnarray} \label{n_thermal} 
	P_n=N_b h f B_{\mathrm{det}}.
\end{eqnarray}
{\color{black}In this equation,} $N_b=[\exp(hf/k_BT)-1]^{-1}$ is the mean photon number of the background thermal noise per mode, $f$ denotes the signal frequency. Moreover, $h=6.63\times10^{-34}~\mathrm{j.s}$ is the Planck constant, $k_B=1.38\times10^{-23}~\mathrm{j.K}$ is the Boltzmann constant, and $T$ {\color{black}denotes the environmental} temperature. For instance, for $f=10 ~\mathrm{GHz}$, $T=300~ \mathrm{K}$, and $B_{\mathrm{det}}=200~\mathrm{kHz}$, we have $N_b\simeq624$ and $P_n\simeq-120.82~\mathrm{dBm}$.

Equation \eqref{max_range_equation} can be analytically solved to obtain an explicit expression for $R_{\mathrm{max}} $. The solution of Eq.~\eqref{max_range_equation} can be {\color{black}represented} in term of the Lambert W-function as follows \cite{corless1996lambert, steinvall2008laser}
\begin{eqnarray} \label{rangeEqDirectFinal}
	R_{\mathrm{max}}^{\mathrm{(DDR)}} =  \frac{20}{\rm{ln(10)} \gamma} W_0\bigg[\frac{\rm ln(10)}{20}\gamma \bigg(\frac{M}{\mathrm{{SNR}_{th}}}\bigg)^{1/4}R_c^{\rm (DDR)}\bigg],
\end{eqnarray}
in which $W_0(x)$ denotes the principal branch of the Lambert's W function, which is a real and increasing function for $x\ge -1/e$. 
This equation {\color{black}determines} the maximum detection range {\color{black}of a} conventional direct detection radar in terms of {\color{black}the} system and environmental parameters. Several key insights can be {\color{black}understood from this expression}: the maximum {\color{black}detection} range can be {\color{black}increased by decreasing} of the atmospheric absorption coefficient $\gamma$, and {\color{black}by} {\color{black}increasing of} the transmitted signal power $P_t$ {\color{black}and the integration samples number} $M$.

Equation~\eqref{rangeEqDirectFinal} {\color{black}demonstrates} that for a given transmit power $P_t$, a critical factor influencing the maximum detection range is the threshold signal-to-noise ratio required for target detection. 
This threshold depends on the sensitivity of the microwave {\color{black}detector} and the specific detection strategy employed in the radar receiver. 
Conventional direct-detection radars typically operate with $\mathrm{SNR}_{\mathrm{th}}$ values in the range of $10\text{--}20~\mathrm{dB}$ \cite{skolnikBook}.  
By utilizing recently developed high-efficiency SMPDs at the radar receiver, $\mathrm{SNR}_{\mathrm{th}}$ can be significantly reduced. 
Analogous to the well-known single-photon LiDARs, we refer to this class of direct-detection radars {\color{black}that} {\color{black}utilize} SMPDs at {\color{black}their} receivers as single-photon radars (SPRs). This improvement and its implications will be discussed in the following sections.

\subsection{ The Single-Photon Radar {\color{black}Idea}} \label{subsec2-2_SPMR}
\begin{table*}
	\centering
	\caption{Summary of selected single microwave-photon detectors demonstrated to date. }
	\begin{tabular}{ccccccccc}
		\toprule
		&
		Frequency & 
		Efficiency& 
		Dark count rate / Probability& 
		Bandwidth &
		Dead time &
		Sensitivity& 
		Year \\
		&(GHz) & ($\%$) & (cps) / - & (MHz) & (ns) & ($\rm W/\sqrt{\rm Hz}$) & \\
		\toprule
		\cite{pallegoix2025enhancing} & 7 & 80 & 30 cps & 0.1-1 & - & $3 \times 10^{-23}$ &2025 \\
		\cite{balembois2024cyclically} & 7 & 43 & 85 cps & 1 & - & $10^{-22}$  &2024 \\
		\cite{dassonneville2020number} & 10.220 & 96 & $0.030 \pm 0.002$ & 20 & 4500 & - &2020  \\
		\cite{inomata2016single}  & 10 & 66 & $0.014\pm 0.001$ & 16 & 400 & - &2016 \\
		\cite{oppliger2025high} & $3-5.2$ & 70 & - & - & 3  & $5 \times 10^{-19}$ &2025   \\ 
		\cite{petrovnin2024microwave} & 6.042 & 73 & 167 kcps & 0.7 & 3000 & $3.228\times10^{-21}$ & 2024 \\
		\toprule
	\end{tabular}
	\label{tab01}
\end{table*}

\begin{figure*}[!htb]
	\centering
	\begin{subfigure}[b]{0.47\textwidth}
		\includegraphics[width=\textwidth]{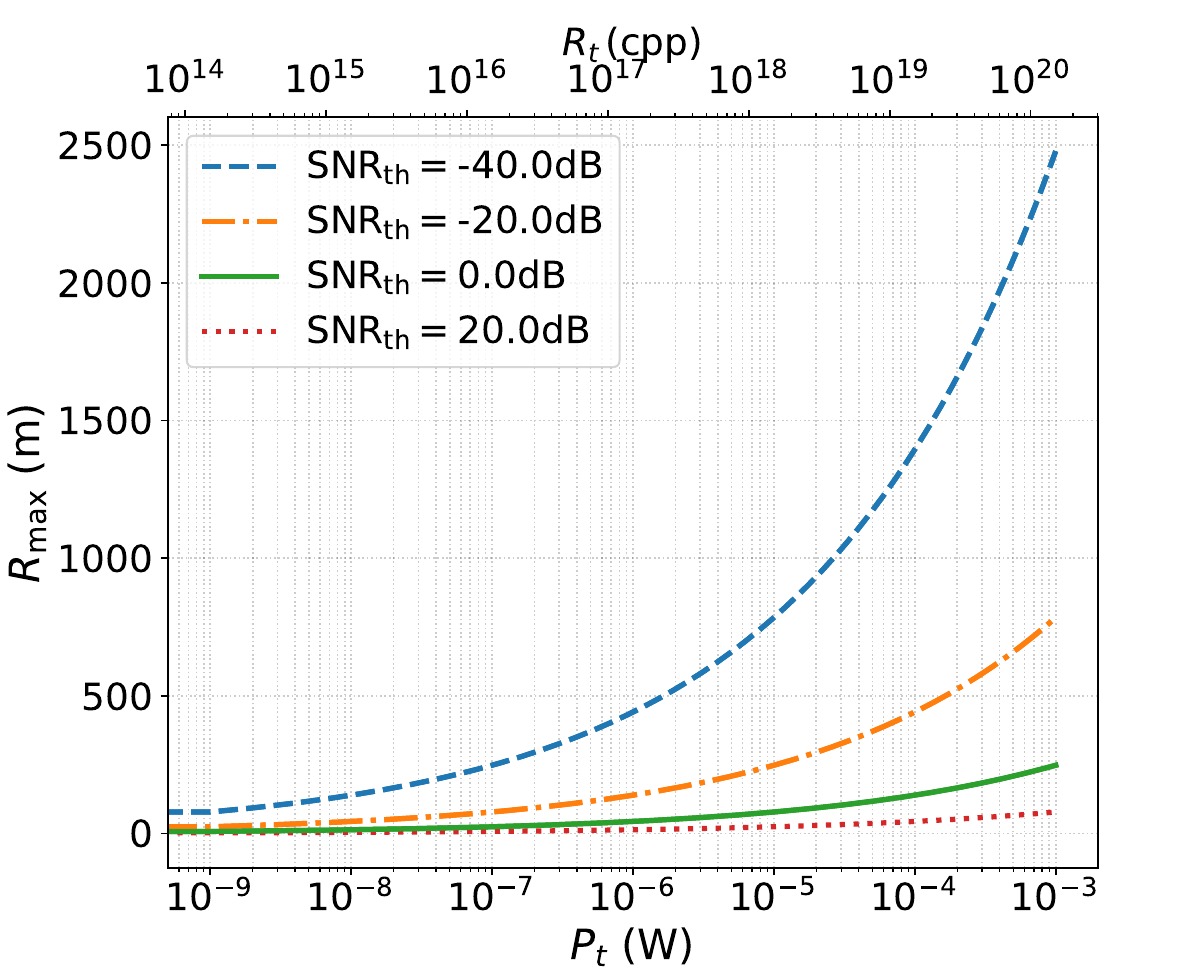}
		\caption{}
		\label{fig2a}
	\end{subfigure}
	\hfill
	\begin{subfigure}[b]{0.47\textwidth}
		\includegraphics[width=\textwidth]{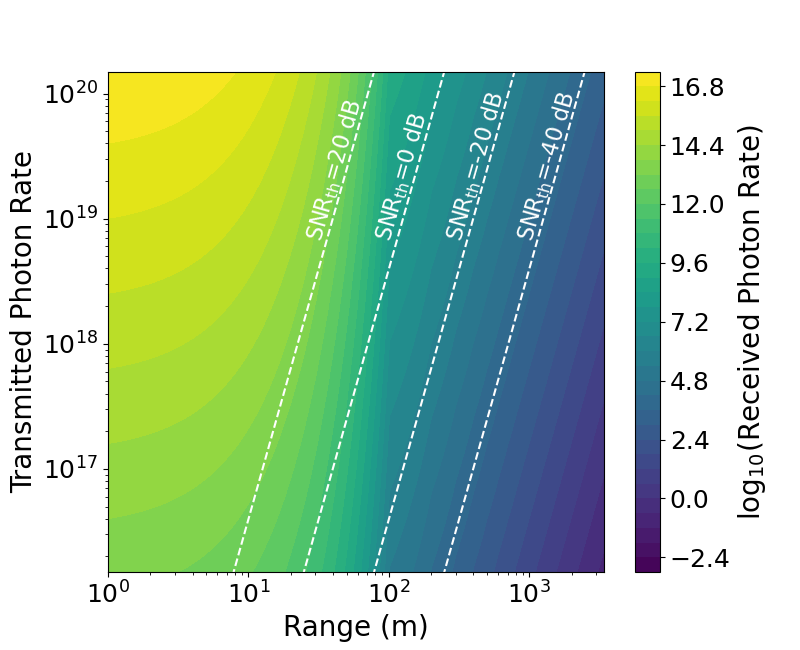} 
		\caption{}
		\label{fig2b}
	\end{subfigure}
	\caption{(color online) (a) {\color{black}The m}aximum detection range of {\color{black}the} direct detection radar {\color{black}versus} the transmitted signal power $P_t$ (lower horizontal axis) and its relevant transmitted photon rate $R_t=P_t/(hf)$ (uper horizontal axis) in {\color{black}the} high brightness regime for {\color{black}the} different values of {\color{black}the} threshold SNR as $\rm SNR_{th}=\{ -40, -20, 0 , 20\}dB$. (b) Contour plot describing the photon rate of the received signal {\color{black}versus} the target range $R$ and the photon rate of the transmitted signal $R_t = P_t/hf$. The {\color{black}white-}dashed lines in this figure represents the \textit{minimum} transmitted photon rate required for resolving the target at range $R$ for {\color{black}the fixed values of} $\mathrm{SNR}_\mathrm{th}$. The transmitted signals with powers under these curves {\color{black}are} not sufficient for detecting the target with the considered {\color{black}threshold SNR}. The {\color{black}other system parameters are as} $f=\rm 10~GHz$, $\sigma = 0.5~\rm m^2$, $G = \rm 15~dB$, $\varepsilon_a = 0.5$, $A=0.01\pi \rm m^2$, $T=\rm 300~K$, $\tau_{ \mathrm{int}}=100~\rm ms$, $B_{\rm det}=200~\rm kHz$, and $\gamma = 0.007 \rm dB/km$ \cite{ITUR}.
	}
	\label{fig2}
\end{figure*}

As demonstrated previously, the threshold SNR of the radar receiver plays a critical role in determining the maximum detection range of a direct-detection radar.
The development of SMPDs, which offer enhanced sensitivity, {\color{black}provides a substantial reduction of} $\mathrm{SNR}_{\mathrm{th}}$.
This, in turn, {\color{black}extends} the operational range of conventional direct-detection radars {\color{black}substantially}. {\color{black}Note that this is important} in scenarios where traditional {\color{black}radars} {\color{black}have a limited range} {\color{black}due to low transmit power. In these cases, SPRs} can {\color{black}operate} effectively as covert radars. {\color{black}The idea of the SPR is similar to single-photon LiDAR, which utilizes single optical-photon detectors (SOPDs) in the LiDAR receiver \cite{liDAR200km2021}}.
While {\color{black}SMPDs} are not yet commercially available, substantial {\color{black}progress has} been made toward their development. {\color{black}Existing implementations of SMPDs} have so far been demonstrated only as {\color{black}a} proof-of-concept in laboratory environments \cite{pallegoix2025enhancing, balembois2024cyclically, dassonneville2020number, inomata2016single, oppliger2025high, petrovnin2024microwave}.

Despite their promise, implementing single-photon detection in the GHz frequency range is technically challenging due to {\color{black}the fact that} photon energy is approximately five orders of magnitude lower than that of optical photons. However, in {\color{black}Ref.}~\cite{inomata2016single}, {\color{black}an} SMPD {\color{black}is} implemented using an impedance-matched artificial $\Lambda$ system comprising the dressed states of a driven superconducting qubit coupled to a microwave resonator. This SMPD achieved a high single microwave-photon detection efficiency of $0.66 \pm 0.06$, with a low dark-count probability of $0.014 \pm 0.001$ at $\sim 10~\mathrm{GHz}$ frequency.
Also, in {\color{black}Ref.}~\cite{oppliger2025high} an SMPD with an efficiency of 70\% in the frequency range {\color{black}of} $3$--$5.2~\mathrm{GHz}$ {\color{black}is} realized based on a hybrid system comprising {\color{black}a} double quantum dot charge qubit electrostatically defined in a GaAs/AlGaAs heterostructure, coupled to a high-impedance Josephson junction array cavity.
{\color{black}Moreover, in Ref.}~\cite{pallegoix2025enhancing}, {\color{black}an} SMPD {\color{black}based on {\color{black}a} superconducting transmon qubit} {\color{black}with a detection} efficiency of 0.8 at {\color{black}the} frequency of $7~\mathrm{GHz}$, {\color{black}and with the} dark-count rate (DCR) below 30 counts per second (cps) {\color{black}is reported}.
{\color{black}In addition, in Ref.}~\cite{balembois2024cyclically}, a bolometric SMPD based on {\color{black}the} irreversible photon absorption by {\color{black}a} transmon qubit via a four-wave-mixing {\color{black}process} is demonstrated {\color{black}with} quantum efficiency about 45\% near {\color{black} the frequency of} $7~\mathrm{GHz}$.
The features of some {\color{black}implemented} SMPDs {\color{black}are} summarized in Table~\ref{tab01}.
It is worth noting that the sensitivity of classical receiver antennas is in the range of $10^{-13}$--$10^{-18}~\mathrm{W}/\sqrt{\mathrm{Hz}}$, which is about one to five orders of magnitude greater than SMPDs.

{\color{black}The technological {\color{black}advances} in {\color{black}single} microwave-photon detection {\color{black}suggest} a novel class of quantum direct-detection radars as single-photon microwave radars, analogous to single-photon terahertz radars and LiDARs \cite{liu2021target, liDAR200km2021}.}
An important aspect of this analogy is that all signal-processing techniques previously developed for single-photon LiDARs, such as time-correlated single-photon counting (TCSPC) or photon-statistical analysis \cite{hirvonen2016wide}, are now applicable to single-photon microwave radars. {\color{black}These processing techniques} can {\color{black}efficiently} reduce the threshold SNR required for target detection, leading to radar-signal identification at SNR levels as low as approximately $-40~\mathrm{dB}$, {\color{black}{\color{black}whereas} it is about $10\text{--}20~\mathrm{dB}$ for classical direct-detection radar systems}.

To illustrate the impact of the threshold SNR, in Fig.~\ref{fig2a} we {\color{black}have plotted} the maximum detection range of a direct-detection radar, $R_{\rm max}$, as a function of the transmitted {\color{black}signal} power, $P_t$, and the corresponding photon rate, $R_t = P_t / hf$, for various threshold SNR from $-40~\mathrm{dB}$ to $+20~\mathrm{dB}$.
{\color{black}In this figure, the system parameters are $f=\rm 10~GHz$, $\sigma = 0.5~\rm m^2$, $G = \rm 15~dB$, $\varepsilon_a = 0.5$, $A=0.01\pi \rm m^2$, $T=\rm 300~K$, $\tau_{ \mathrm{int}}=100~\rm ms$, $B_{\rm det}=200~\rm kHz$, and $\gamma = 0.007 \rm dB/km$ \cite{ITUR}}.
As expected, {\color{black}the maximum detection range} increases {\color{black}by} increasing {\color{black}the} transmit power. However, for a fixed $P_t$, the maximum detection range also increases {\color{black}by decreasing} the threshold SNR.
For example, with a transmit power of $P_t = 1~\mathrm{nW}$, {\color{black}which} corresponds to {\color{black}the} photon rate of $R_t = 1.5 \times 10^{14}~\mathrm{cpp}$ (counts per pulse), the detection range is approximately $2.9~\mathrm{m}$ at $\mathrm{SNR}_{\mathrm{th}} = 20~\mathrm{dB}$ (dotted red line).
{\color{black}However,} this range increases to nearly $80~\mathrm{m}$ when the threshold SNR is reduced to $-40~\mathrm{dB}$ (dashed red line). In the high-brightness regime, with $P_t = 1~\mathrm{mW}$, {\color{black}which} corresponds to a photon rate of $R_t = 1.5 \times 10^{20}~\mathrm{cpp}$, the maximum detection range is approximately $79.5~\mathrm{m}$ for $\mathrm{SNR}_{\mathrm{th}} = 20~\mathrm{dB}$, {\color{black}while {\color{black}it} is} about $2480~\mathrm{m}$ when the threshold SNR is decreased to $-40~\mathrm{dB}$.

One {\color{black}critical} consideration in the operation of single-photon microwave radars is whether a sufficient number of signal photons {\color{black}returned} to the receiver for detection or not. {\color{black}{\color{black}It}  is important especially when the power of the transmitted signal is extremely weak}.
To address this, we compare the power (or {\color{black}the} photon rate) of the received signal with the minimum detectable signal (MDS), {\color{black}which} is given by
\begin{equation} \label{MDS_formula}
	P_{\rm MDS} = \frac{1}{M} \mathrm{SNR}_{\mathrm{th}} \times P_n,
\end{equation}
where $M$ is the number of integration samples, and $P_n$ is the noise power. {\color{black}Note that the} detection occurs only if $P_r \geq P_{\rm MDS}$, {\color{black}which is} consistent with the detection criterion described in Sec.~\ref{subsec2-1_review}. 
{\color{black}By} substituting the expression for {\color{black}the power of the received signal, $P_r$,} from Eq.~\eqref{power1}, into this inequality, we obtain
\begin{equation} \label{eq_min_transmit_power}
	P_t \geq \frac{\mathrm{SNR}_{\mathrm{th}} \cdot P_n}{M \cdot \eta(R)}.
\end{equation}
{\color{black}This relation} defines the minimum power (or {\color{black}the} photon rate) {\color{black}of  the transmitted signal for which the power of the} received signal exceeds the MDS. 

Figure~\ref{fig2b} shows the received photon rate, $R_r = P_r / hf$, as {\color{black}a} function of the transmitted photon rate, $R_t = P_t / hf$, and the target range $R$. In this figure, the {\color{black}white-}dashed lines indicate the \textit{minimum} transmitted photon rate {\color{black}that is} required to detect {\color{black}the} target at {\color{black}given} range for each value of $\mathrm{SNR}_{\mathrm{th}}$. If the transmitted photon rate falls below the respective threshold curve, the photon rate of the received signal will be less than the MDS, and {\color{black}thus, the detection of target} is no longer possible.
Also, this figure shows that for $\mathrm{{SNR}_{th}}=20~\rm dB$,  when the transmitter emits signal photons with the rate of $1.5\times10^{20}~\rm cpp$ (equivalent to about $\sim 1~\rm mW$), the system is able to identify targets at ranges up to $77~\rm m$. {\color{black}In this case, the photon rate of the received signal} is approximately $\sim 10^{12}~\rm cpp$. {\color{black}Moreover, for} $\mathrm{{SNR}_{th}}=-40~\rm dB$, the transmitted signal photon rate should be at least $4\times10^{18}~\rm cpp$ (equivalent to $~\rm 15~\mu\rm W$) {\color{black}in order to} detect targets {\color{black}at the} range of $1~\rm km$. Under this condition, {\color{black}the photon rate of the} received signal is greater than $\sim 10^{3.6}~\rm cpp$.


\section{Quantum-Entangled Noise Radars} \label{sec3.NoiseRdar}

In direct-detection radars, the returned signal can be easily masked by a noise source or unwanted {\color{black}signals} {\color{black}incident on} the receiver. {\color{black}To overcome this limitation, it is proposed to} extract the desired signal from {\color{black}the} noise by correlating the {\color{black}received} signal with the reference signal, which is a copy of the transmitted signal. \cite{narayananNoiseRadarDesign,lukinNoiseRadarTechnology,noiseRadarOverview,efficientProcessingNoiseRadar2018}. This is the main operational principle of noise radar systems.

\begin{figure}[!htb]
	\centering
	\includegraphics[width=250pt]{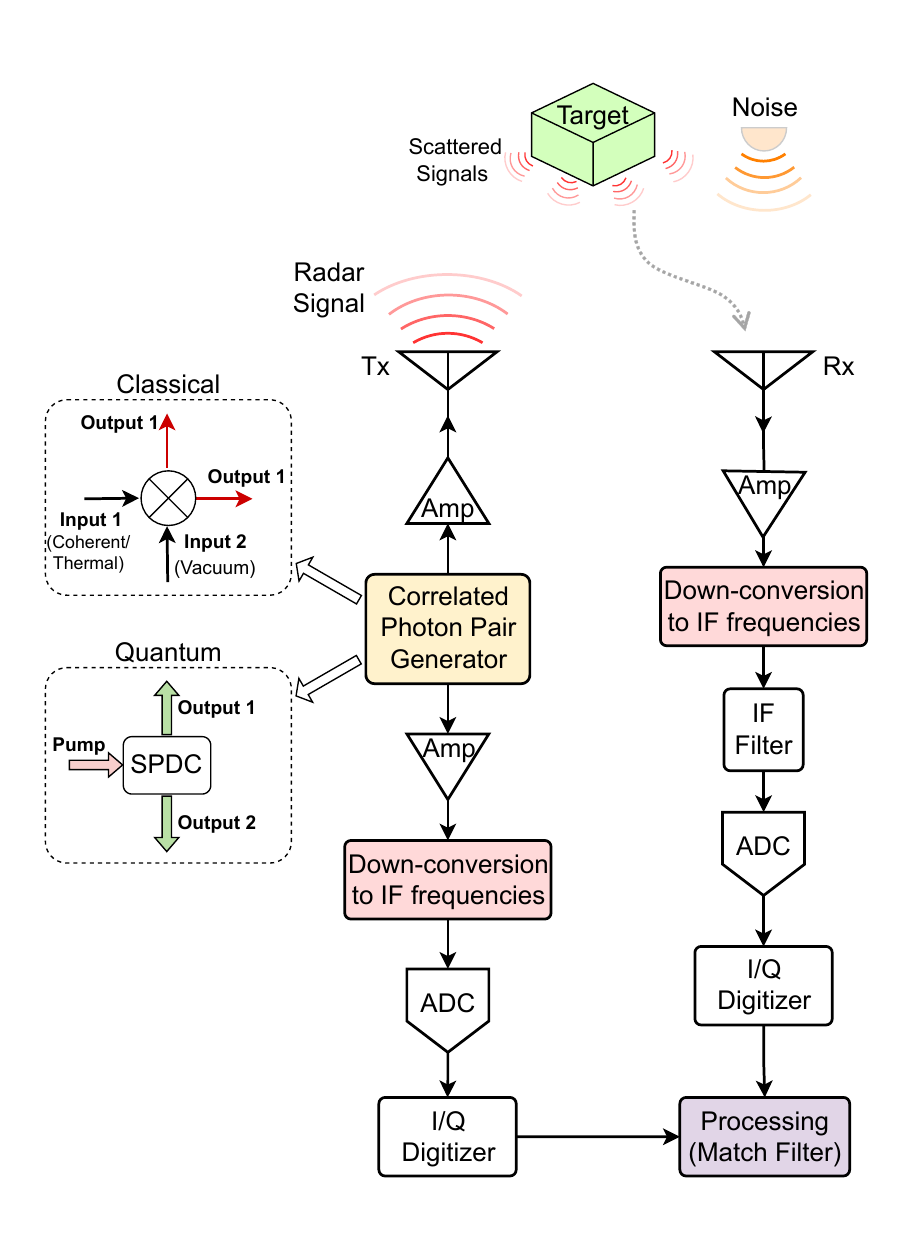}
	\caption{The schematic diagram of a mono-static noise radar system. Amp.: Amplifier; ADC: Analog to digital converter; SPDC: Spontaneous parametric down-conversion.}
	\label{fig3}
\end{figure}

The simplified schematic diagram of a noise radar is shown in Fig.~\ref{fig3}. In this system, a pair of correlated fields, either classical or quantum, is generated by the microwave source. One of these fields, referred to as the signal, is amplified and {\color{black}then} transmitted toward the target region. {\color{black}However,} the other one, known as the idler, is amplified, detected, and {\color{black}recorded} as {\color{black}the} reference. This reference is later used to distinguish the desired signal from unwanted noise in the received field {\color{black}via correlation}.
At the receiver, the {\color{black}received} signal, which may include the back-scattered {\color{black}signal} (if {\color{black}the} target is {\color{black}present}) along with various noise contributions{, \color{black}is detected}. {\color{black}It is then} amplified and down-converted to {\color{black}an} IF frequency {\color{black}in the radar base-band}. {\color{black}After that, it is} filtered and digitized using {\color{black}a high-sample-rate, high-bandwidth} analog-to-digital converter (ADC). {\color{black}The} target information is then extracted through a process known as matched filtering \cite{balajiMythToReality2020}.

{It is important to note that a key difference between the classical and quantum noise radars is {\color{black}their} microwave source used to generating correlated signal-idler fields.}
In the classical case, the signal and idler are classically correlated electromagnetic (EM) fields. In contrast, in the quantum case, the signal and idler are entangled EM waveforms, meaning that the corresponding quantum state shows strong non-classical correlations \cite{balajiMythToReality2020,barzanjehQIlluminationOMS2015,quantumilluminationSethLloyd2008}.

{\color{black}In noise radars,} the degree of correlation between the signal and idler fields is typically quantified using the covariance matrix constructed from the in-phase ($I$) and quadrature ($Q$) voltage components of the fields. This matrix is defined as follows \cite{quantumEnhancedNoiseRadar2019,sorelli2022}
\begin{eqnarray} \label{covarianceMatrix1}
	\mathrm{Cov} = \begin{pmatrix}
		C_\mathrm{SS} & C_\mathrm{SI} \\
		C_\mathrm{IS} & C_\mathrm{II}
	\end{pmatrix},
\end{eqnarray}
{\color{black}where} $C_{nm}$ ($n,m \in \{\mathrm{S}, \mathrm{I}\}$) is a $2 \times 2$ {\color{black}block}, {\color{black}and} given by
\begin{eqnarray} \label{covarianceMatrix1blocks}
	C_{nm} = \begin{matrix}
		\braket{\hat{I}_n \hat{I}_m} & \braket{\hat{I}_n \hat{Q}_m} \\
		\braket{\hat{Q}_n \hat{I}_m} & \braket{\hat{Q}_n \hat{Q}_m}
	\end{matrix}.
\end{eqnarray}
Here, $\braket{\cdot}$ denotes the quantum expectation value, and the operators $\hat{I}_k$ and $\hat{Q}_k$ represent the quantum-mechanical in-phase and quadrature voltage components of the electromagnetic field, {\color{black}respectively}, which defined as
\begin{eqnarray} \label{IQ_operators}
	\hat{I}_k = &  \frac{1}{\sqrt{2}} ( \hat{a}_k + \hat{a}_k^\dagger ), \\
	\hat{Q}_k = & \frac{1}{i\sqrt{2}} ( \hat{a}_k - \hat{a}_k^\dagger ),
\end{eqnarray}
{\color{black}where} $\hat{a}_k$ and $\hat{a}_k^\dagger$ are the annihilation and creation operators, respectively, associated with {\color{black}the} mode $k$, with $k = \mathrm{S}$ for the signal and $k = \mathrm{I}$ for the idler mode.

The off-diagonal blocks of the covariance matrix, namely $C_{\mathrm{SI}}$ and $C_{\mathrm{IS}}$, represent the correlations between the in-phase ($I$) and quadrature ($Q$) voltage components of the signal and idler fields. The strength of this correlation is quantified by the well-known Pearson correlation coefficient, $\rho$, which is defined as
\begin{eqnarray}  \label{ro_1}
	& \rho = \dfrac{\langle {{\hat{I}_{\rm S}}{\hat{I}_{\rm I}}} \rangle}{\sqrt{\braket{\hat{I}_{\rm S}^2} \braket{\hat{I}_{\rm I}^2}} }.
\end{eqnarray}

Conceptually, the role of the Pearson correlation coefficient, $\rho$, in noise radars is analogous to that of {\color{black}the} signal power in direct detection radars. It is the detection {\color{black}function} that indicates the presence of {\color{black}the} target within the interrogation region \cite{dawood2021ReceiverOperating}, as discussed in Sec.~\ref{sec2:direct_detection_radars}. 
Under the target-present hypothesis ($H_1$), a nonzero correlation exists between the received signal and the retained idler, i.e., $\rho \neq 0$. Conversely, when {\color{black}there is} no target ({\color{black}hypothesis} $H_0$), the correlation {\color{black}disappears}, {\color{black}i.e.,$\rho = 0$}.

To calculate the correlation coefficient between the \textit{detected} signal (at the receiver) and idler fields quantum mechanically using Eq.~\eqref{ro_1}, it is necessary to determine the annihilation operators corresponding to the detected signal and idler fields in terms of the system and environmental parameters. For the signal field, the {\color{black}annihilation operator of} the detected field depends on the presence of the target, {\color{black}which} is given by
\begin{eqnarray}  \label{a_hat_S_det}
	& {\rm H_1: \quad} \hat{a}_{\rm S,1}^{\rm det} = \sqrt{ \eta{(R)} G_{\rm S} } \hat{a}_{\rm S}+ \hat{a}_{\rm S, add}^{\rm (1)},\\
	& {\rm H_0: \quad} \hat{a}_{\rm S,0}^{\rm det} = \hat{a}_{\rm S, add}^{(0)}.
\end{eqnarray}
{\color{black}Here,} $\hat{a}_{\rm S, add}^{(1)}$ and $\hat{a}_{\rm S, add}^{(0)}$ {\color{black}refer} to the annihilation operator of the {\color{black}noise added} to the signal {\color{black}that} reaches to the receiver under $\rm H_1$ and $\rm H_0$ {\color{black}hypotheses, respectively,} {\color{black}which are given by}
\begin{eqnarray}  \label{a_hat_S_add}
	\hat{a}_{\rm S, add}^{(1)} =&  \sqrt{\big[1-\eta(R)\big]G_{\rm S, rec} } \, \hat{a}_{\rm env} + \sqrt{\eta(R) G_{\rm S, rec} }  \hat{L}_{\rm S,tr}^\dagger +\hat{L}_{\rm S, rec}^\dagger , \nonumber \\
	& \\
	\hat{a}_{\rm S, add}^{(0)} =& \sqrt{G_{\rm S, rec}} \hat{a}_{\rm env} + \hat{L}_{\rm S, rec}^\dagger.
\end{eqnarray}
{\color{black}In these relations,} $G_{\rm S}=G_{\rm S, rec}G_{\rm S,tr}$ is the total amplification gain of the signal, $G_{\rm S, tr}$ and $G_{\rm S, rec}$ {\color{black}denote} the signal {\color{black}gains} at the transmitter and receiver, respectively. 
{\color{black}Moreover,} $\hat{a}_{\rm env}$ gives the annihilation operator of the thermal background noise. 
{\color{black}In addition,} $\hat{L}_{\rm S, tr}$ and $\hat{L}_{\rm S, rec}$ denote the Langevin noise operators that {\color{black}describe} the {\color{black}noise added} to the fields {\color{black}through} the RF amplifiers at the transmitter and receiver, respectively, such that \cite{boyd1994quantum}
\begin{eqnarray} 
	\hat{L}_{\rm S, tr}=&\sqrt{G_{\rm S, tr}-1} \hat{c}_{\rm S,tr},   \\
	\hat{L}_{\rm S, rec}=&\sqrt{G_{\rm S, rec} -1} \hat{c}_{\rm S,rec}.
\end{eqnarray}
{\color{black}Here,} $\hat{c}_{\rm S,tr}$ and $\hat{c}_{\rm S,rec}$ are bosonic field operators satisfying {\color{black}the} commutation relation $[ \hat{c}_{k} , \hat{c}_k^\dagger]=1 $. {\color{black}It should be noted} that {\color{black}the mean photon number of the {\color{black}noise added} through the RF amplifier is given by}
\begin{equation}\label{N_k}
	N_k^{\rm amp} \equiv \braket{\hat{c}_k^\dagger \hat{c}_k}=k_B T_{\mathrm{eff}, k}(10^{\mathrm{NF}_k/10}-1)/hf_k
\end{equation}
in which $T_{\mathrm{eff}, k}$ is the effective temperature of the amplifier, $\mathrm{NF}_k$ denotes its noise figure in dB scale, $f_k$ is the frequency, and $ k\in \{\rm (S, rec); (S,tr); I\} $ is dummy index.

The annihilation operator of the detected idler mode is independent of the target presence, and given by 
\begin{eqnarray}  \label{a_hat_I}
	& \hat{a}_{\rm I}^{\rm det} = \sqrt{G_{\rm I}} \hat{a}_{\rm I} +  \hat{L}_{\rm I}^\dagger,
\end{eqnarray}
in which $G_{\rm I}$ is the Idler amplification gain, and $\hat{L}_{\rm I}=\sqrt{G_{\rm I}-1}\hat{c}_{\rm I}$.

The radar range equation for noise radars should establish a relationship between the correlation coefficient, $\rho$, and the target range, $R$. 
In the following, we present a comprehensive and explicit formulation of the range equation for both classical and quantum noise radars.

\subsection{ Maximum Detection Range of Quantum-Entangled Noise Radar} \label{Sub.Sec3-1_QTMS_radar}

\begin{figure*}[!htb]
	\centering
	\begin{subfigure}[b]{0.47\textwidth}
		\includegraphics[width=\textwidth]{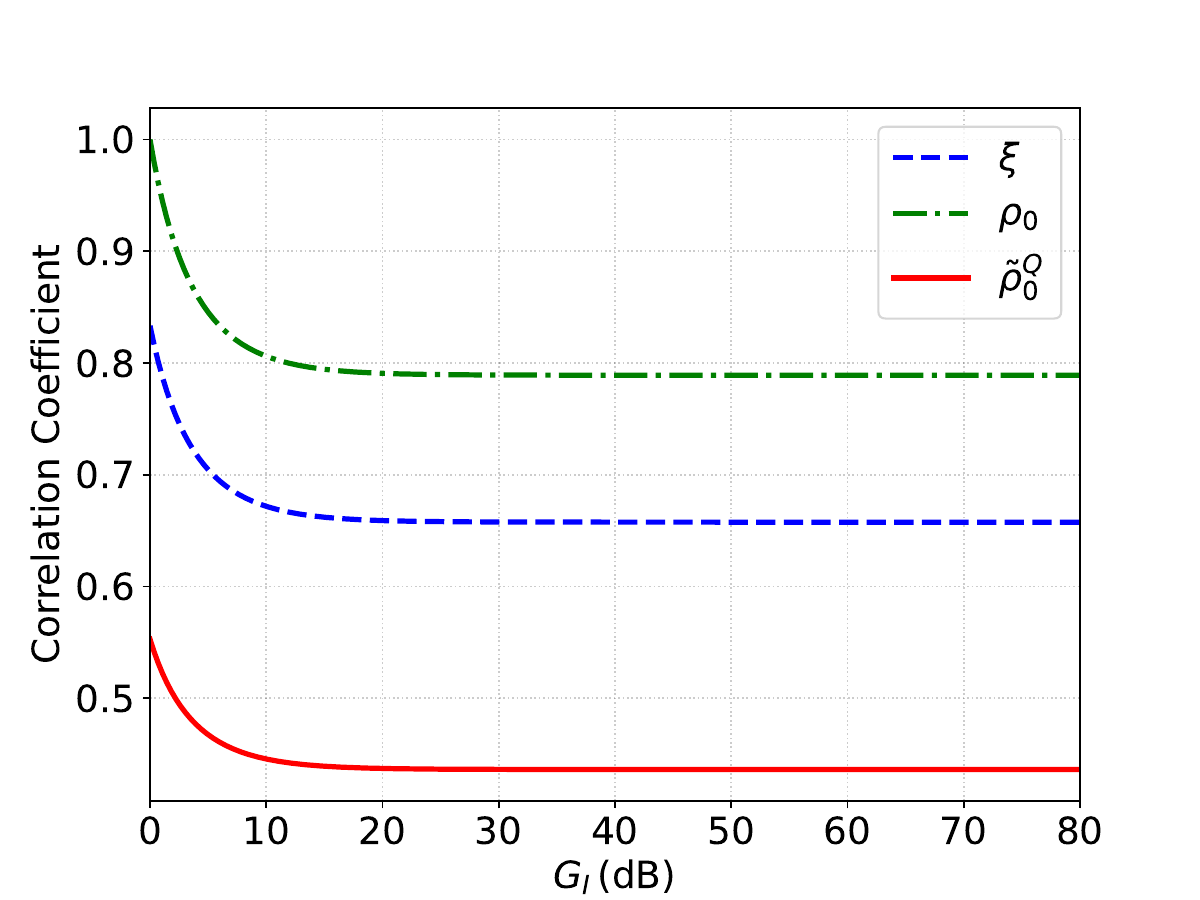}
		\caption{}
		\label{fig4a}
	\end{subfigure}
	\hfill
	\begin{subfigure}[b]{0.47\textwidth}
		\includegraphics[width=\textwidth]{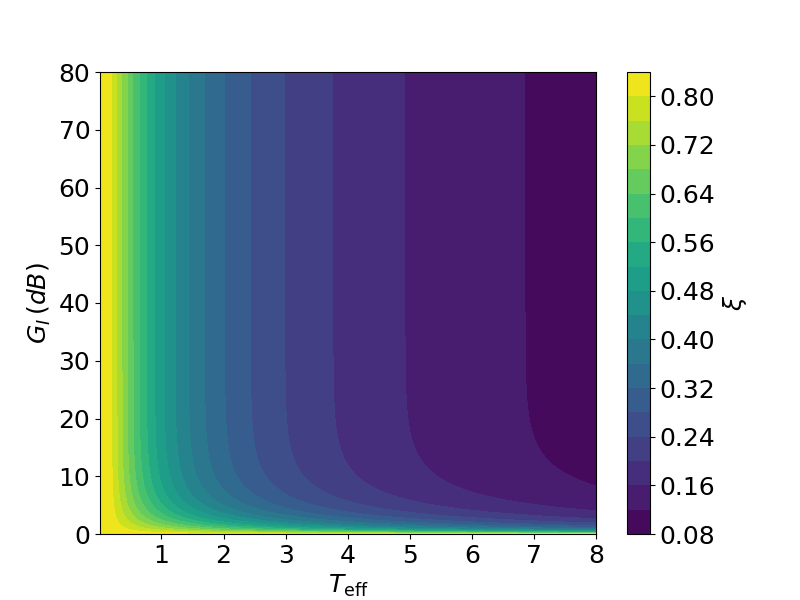} 
		\caption{}
		\label{fig4b}
	\end{subfigure}
	\caption{(Color online) (a) {\color{black}The} modified correlation coefficient $\tilde{\rho}_0$, the correlation coefficient $\tilde{\rho}_0$ reported in {\color{black}Ref} \cite{balaji2022PerformancePrediction}, and the correction term $\xi$ {\color{black}versus} the amplification gain of the idler mode $G_{\mathrm{I}}$. The noise figure and {\color{black}the} effective temperature of the idler amplifier are fixed at $\mathrm{NF} = 1\,\mathrm{dB}$ and $T_{\mathrm{eff}} = 500\,\mathrm{mK}$, respectively.
		(b) The correction term $\xi$ as {\color{black}the} function of the idler amplifier temperature $T_{\mathrm{eff}}$ and gain $G_{\mathrm{I}}$ for {\color{black}the} noise figure of $\mathrm{NF} = 1\,\mathrm{dB}$. In both figures, the {\color{black}photon number per mode} is set to $N_s = 0.1$.
	}
	\label{fig4}
\end{figure*}

In quantum-entangled {\color{black}noise} radars, a microwave quantum two-mode squeezed vacuum {\color{black}(TMSV)} state of the electromagnetic field (see Appendix.~\ref{appendix_squeezing}) is employed as the quantum-correlated signal-idler pairs. In the number-state basis, the TMSV state is given by \cite{guha2009gaussian, frasca2020EntangledCoherentQRadar}
\begin{eqnarray}  \label{psi_TMSV}
	\ket{\psi}_{\rm TMSV} = \sum_{n=0}^{\infty}{ \sqrt{\frac{N_s^n}{(N_s+1)^{n+1}}} \ket{n}_{\rm S} \ket{n}_{\rm I} },
\end{eqnarray}
{\color{black}in which} $N_s$ is the average number of photons per mode, {\color{black}that relates} to the transmitted power by $P = hfN_sB$, with $f$ and $B$ denoting the field frequency and the source bandwidth, respectively. 
Under the hypothesis $\mathrm{H}_1$, the quantum state of the received signal and idler pair is described by the density matrix $\hat{\rho} = \hat{\rho}_{\rm SI} \otimes \hat{\rho}_{\rm n,S} \otimes \hat{\rho}_{\rm n,I}$, such that $\hat{\rho}_{\rm SI} = \ket{\psi_{\rm SI}}\bra{\psi_{\rm SI}}$ represents the density matrix of the correlated signal-idler pair, and $\hat{\rho}_{\mathrm{n}, k}$ refers to the density matrices of the {\color{black}noise added} to the mode $k\in \{\rm S, I\}$. 
Using this expression, {\color{black}one can calculate} the terms appeared in the Pearson correlation coefficient (Eq.~\ref{ro_1}) {\color{black}as follows}
\begin{eqnarray}
	\braket{\hat{I}_{\rm S}^2}_{\rm QI} &\simeq \eta(R) G_{\rm S} \bigg(N_s + \frac{1}{2}\bigg) + N_{\rm S, add}^{(1)},\label{Pearson_factors1}\\ 
	\braket{\hat{I}_{\rm I}^2}_{\rm QI} &\simeq G_{\rm I} \bigg(N_s + \frac{1}{2}\bigg) + N_{\rm I, add},\label{Pearson_factors2}\\ 
	\braket{\hat{I}_{\rm S}\hat{I}_{\rm I}}_{\rm QI} &= \sqrt{\eta(R) G_{\rm S} G_{\rm I} N_s(N_s + 1)}. \label{Pearson_factor3}
\end{eqnarray}
{\color{black}{\color{black}In these relations,} $N_{\rm S, add}^{(1)}$ and $N_{\rm I, add}$ represent the mean photon number of the {\color{black}noise added} to the signal and idler modes, respectively, and assumed to be very large compared to the zero-point fluctuations in the signal and idler noise modes.} 
Using Eqs.~\ref{a_hat_S_add} and \ref{a_hat_I}, one can calculate the mean photon number of the {\color{black}added} noise under the hypothesis $\mathrm{H}_1$ as follows
\begin{eqnarray}  \label{added_nooise_2}
	N_{\rm S, add}^{\rm (1)} =&  \big[ 1-\eta(R) \big] G_{\rm S,rec} N_{\rm env} + (G_{\rm S,rec}-1)(N_{\rm S, rec}^{\rm amp}+1) \notag\\
	&+\eta(R)G_{\rm S} (G_{\rm S,tr}-1) (N_{\rm S, tr}^{\rm amp}+1) ,\\
	N_{\rm I, add}=&(G_{\rm I}-1)(N_{\rm I}^{\rm amp}+1). 
\end{eqnarray}
Assuming that the RF amplifier in the radar transmitter affects {\color{black}the} signal and idler modes equally (i.e., $G_{\rm S,tr} \approx G_I$ and $N_{\rm S,tr}^{\rm amp}\approx N_{\rm I}^{\rm amp}$),  and $\eta(R) \ll 1$ (which is the case of very {\color{black}low-reflecting} objects in long ranges), {\color{black}one can rewrite} $N_{\rm S, add}^{\rm (1)}$ as {\color{black}follows}
\begin{eqnarray}  \label{added_nooise_3}
	& N_{\rm S, add}^{\rm (1)} = \eta(R) G_{\rm S, rec} N_{\rm I, add} + N_n ,
\end{eqnarray}
in which
\begin{eqnarray}  \label{added_nooise_4}
	& N_n \simeq G_{\rm S, rec} N_{\rm env} + (G_{\rm S, rec}-1)(N_{\rm S, rec}^{\rm amp} + 1).
\end{eqnarray}
{\color{black}In these relations,} $N_{\rm env}=[\exp(hf/k_B T_{\rm env})-1]^{-1}$ being the mean photon number of the environmental thermal noise.  
By substituting the results obtained in Eqs.~\eqref{Pearson_factors1}-\eqref{Pearson_factor3} into Eq.~\eqref{ro_1}, {\color{black}after some algebraic calculations}, one can obtain the Pearson correlation coefficient {\color{black}as follows}
\begin{eqnarray}  \label{Pearson_QI}
	& \rho_{\rm QI} (R) = \frac{ \tilde{\rho}_0^{\rm QI} }{\sqrt{ 1+ \frac{1}{\eta(R) G_{\rm S, rec}} \frac{N_n}{\braket{\hat{I}_{\rm I}^2}} } }. 
\end{eqnarray}
{\color{black}Here,} $\tilde{\rho}_0^{Q}$ represents the modified Pearson correlation coefficient between the signal and idler modes at the source  (i.e., at $R=0$), {\color{black}and} is given by
\begin{eqnarray} \label{quantum_Pearson}
	& \tilde{\rho}_0^{\rm QI}=\sqrt{\rho_0^2 - \xi^2},
\end{eqnarray}
with
\begin{eqnarray} 
	&&  \rho_0 = 1 - N_{\rm I, add}/\braket{\hat{I}_{\rm I}^2}, \label{rho_0}\\
	&&   \xi = G_{\rm I}/2\braket{\hat{I}_{\rm I}^2}. \label{xi}
\end{eqnarray}
Equation~\eqref{Pearson_QI} {\color{black}is} the range equation for the quantum-entangled {\color{black}noise} radar, which {\color{black}enables us to evaluate} the residual Pearson correlation coefficient between the returned signal (from the target at {\color{black}a range $R$}) and idler modes.

\begin{figure*}[!htb]
	\centering
	\begin{subfigure}[b]{0.47\textwidth}
		\includegraphics[width=\textwidth]{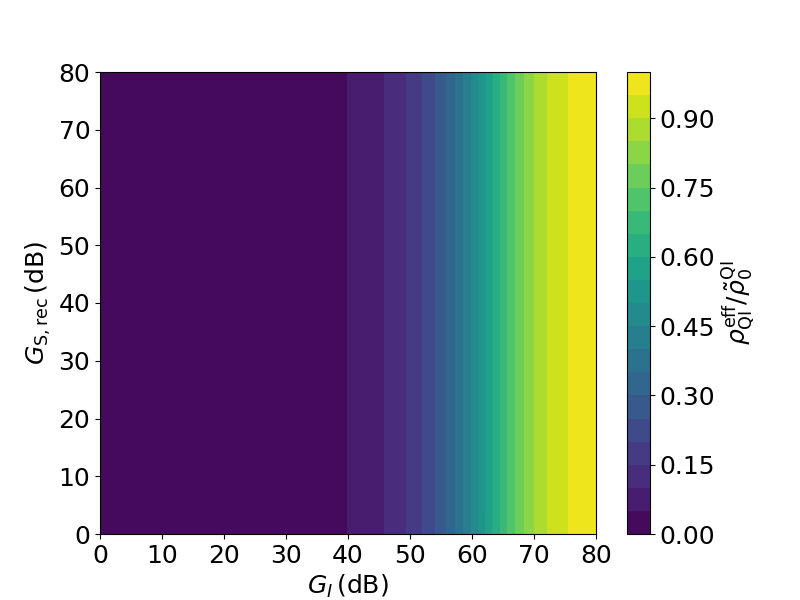}
		\caption{}
		\label{fig5a}
	\end{subfigure}
	\hfill
	\begin{subfigure}[b]{0.47\textwidth}
		\includegraphics[width=\textwidth]{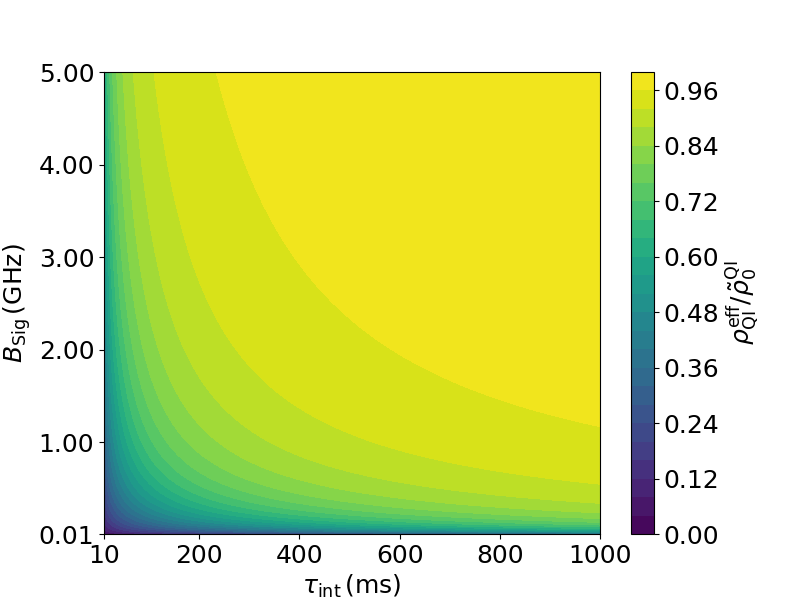} 
		\caption{}
		\label{fig5b}
	\end{subfigure}
	\caption{(Color online) The normalized Pearson correlation coefficient $\rho_Q^\mathrm{eff}/\rho_0^Q$ {\color{black}versus} (a) amplification gain of the idler mode $G_{\rm I}$ and gain of the signal mode at the receiver $G_\mathrm{S, rec}$ for {\color{black}the} target at range of $R=1~\rm km$,
		and (b) {\color{black}the} integration time $\tau_{\rm int}$ and the bandwidth of the signal mode $B_{\rm sig}$ for $G_{\rm I}= G_\mathrm{S, rec} = 80~\rm dB$.
		In both cases, the noise figure and effective temperature of the idler and signal amplifiers are fixed at $\mathrm{NF} = 1\,\mathrm{dB}$ and $T_{\mathrm{eff}} = 100\,\mathrm{mK}$, respectively. {\color{black}Moreover,} the target range is considered to be $R=1~\rm km$, and the {\color{black}photon number per mode} is set to $N_s = 0.1$. Other system parameters are {\color{black}the} same as {\color{black} in Fig.~\ref{fig2}}. 
	}
	\label{fig5}
\end{figure*}

It is important to highlight that Eq.~\eqref{Pearson_QI} represents {\color{black}the} modified version of the expression introduced in Ref~\cite{balaji2022PerformancePrediction}. {\color{black}It contains} an additional correction term, $\xi^2$, which depends on the amplifier gain, {\color{black}the} added noise, and the average photon number per mode. {\color{black}This correction term} is generally non-negligible and results in a reduction of the correlation {\color{black}coefficient} {\color{black}compared} to {\color{black}the value} {\color{black}reported in Ref.~\cite{balaji2022PerformancePrediction}.
To show this, we have illustrated the} behavior of the modified Pearson correlation coefficient $\tilde{\rho}_0^{\rm QI}$, the correction term $\xi$, and the original correlation coefficient $\rho_0$ versus the idler amplification gain $G_{\rm I}$ in Figure~\ref{fig4a} {\color{black}for} $N_s = 0.1$, $\mathrm{NF} = 1~\mathrm{dB}$, and $T_{\rm eff} = 500~\mathrm{mK}$. As shown in {\color{black}this} figure, for the parameters considered here, the modified correlation coefficient is approximately $44.72\%$ lower than the coefficient $\rho_0$. {\color{black}This}  that the contribution of the amplifier noise {\color{black}must be included} in {\color{black}the} Pearson {\color{black}correlation} coefficient.
From Eq.~\eqref{quantum_Pearson}, it is evident that the modified correlation coefficient $\tilde{\rho}_0^{\rm QI}$ approaches $\rho_0$ in the limit {\color{black}of} $\xi \to 0$.
{\color{black}According} to Eq.~\eqref{xi}, in the regime {\color{black}where} $N_s \ll 1$, {\color{black}in which the QI} protocol outperforms the best classical strategies, this limit is satisfied when $N_{\rm I}^{\rm amp} \gg 1$. {\color{black}On the other hand,} the condition $N_{\rm I}^{\rm amp} \gg 1$ is, in turn, satisfied when the idler amplifier operates at high temperatures.
This behavior is further demonstrated in Fig.~\ref{fig4b}, where the correction term $\xi$ is plotted {\color{black}versus} the idler amplification gain $G_{\rm I}$ and the effective temperature $T_{\rm eff}$. {\color{black}As is} {\color{black}evident}, the correction term is significantly influenced by the effective temperature, while its variation with respect to $G_{\rm I}$ is relatively modest. Specifically, the correction term drops below $0.16$ for $T_{\rm eff} > 5~\mathrm{K}$.
In quantum-entangled {\color{black}noise} radars, due to the fragility of quantum correlations to noise, it is beneficial to employ low-noise amplifiers operating at cryogenic temperatures, for which $N_{\rm I}^{\rm amp}$ is small~\cite{barzanjehExperimentQuantumRadar2020, balaji2019receiver}. Under {\color{black}this condition}, the correction term becomes substantial, and hence, $\tilde{\rho}_0^{\rm QI} < \rho_0$.

If detection occurs by integrating over $M=\tau_{\rm int}B_{\rm det}$ samples, in which $\tau_{\rm int}$ is the integration time and $B_{\rm det}$ is the detection bandwidth, the SNR of the received signal and {\color{black}idler} modes {\color{black}increases} by the factor $M$. Therefore, the effective correlation coefficient {\color{black}becomes}
\begin{eqnarray}  \label{Pearson_QI_eff}
	& \rho_{\rm QI}^{\rm eff} (R) \simeq \frac{ \tilde{\rho}_0^{\rm QI} }{\sqrt{ 1+ \frac{1}{M G_{\rm S,rec} F^2(R)} \big(\frac{R}{R_c}\big)^4 } },
\end{eqnarray}
where $R_c$ {\color{black}represents} the characteristic range of the quantum-entangled {\color{black}noise} radar, {\color{black}and} is given by
\begin{eqnarray}  \label{Characteristic_range_QRadar}
	& R_c = \bigg( \frac{\sigma G A_e \braket{\hat{I}_{\rm I}^2 }_{\rm QI} } {{(4\pi)^2 N_n}} \bigg)^{1/4}.
\end{eqnarray}

In Fig.~\ref{fig5}, we {\color{black}have shown} how key system parameters, such as {\color{black}the signal-idler amplification {\color{black}gains}}, integration time $\tau_{\rm int}$, and signal bandwidth $B_{\rm sig}$, affect the normalized Pearson correlation coefficient, $\rho_\mathrm{QI}^{\rm eff}/\tilde \rho_0^{\rm QI}$. 
As shown in Fig.~\ref{fig5a}, increasing the signal gain at the receiver, {\color{black}$G_{\rm S, rec}$}, has {\color{black}a} {\color{black}negligible} impact on the correlation coefficient, {\color{black}while} increasing the idler gain, $G_{\rm I}$, significantly enhances it. {\color{black}This is due to the fact that} the term {\color{black}$N_n/(G_\mathrm{S, rec} \braket{\hat{I}_{\rm I}^2})$} {\color{black}that appeared in Eq.~\eqref{Pearson_QI}} remains nearly independent of $G_{\rm S, rec}$, but {\color{black}it is scaled} inversely with $G_{\rm I}$.
Moreover, Fig.~\ref{fig5b} shows that increasing either $\tau_{\mathrm{int}}$ or $B_{\rm sig}$ enhances the correlation between the detected signal and idler fields. Increasing both the signal bandwidth and the integration time raises the effective signal power, thereby improving the SNR at the receiver. Note that increasing $B_{\rm sig}$ directly decreases the photon number per mode, $N_s$.

In quantum-entangled {\color{black}noise} radars, the presence of {\color{black}the} target is declared when the effective correlation coefficient exceeds a threshold value, $\rho_{\mathrm{th}}$. This leads to $\rho_{\rm QI}^{\mathrm{eff}}(R) \ge \rho_{\mathrm{th}}$,{\color{black} which is referred {\color{black}to} {\color{black}as} {\color{black}the} detection condition for noise radars}. 
Note that the threshold correlation coefficient, $\rho_{\rm th}$, depends on the number of integration samples $M$ and the false alarm probability $P_{\mathrm{fa}}$ for the target detection, {\color{black}which} {\color{black}is determined by} 
$\rho_{\mathrm{th}} = \sqrt{-\ln P_\mathrm{fa}/M}$
\cite{balaji2019EstimatingCorrelationCoefficients}.
{\color{black}By substituting the effective correlation coefficient obtained in} Eq.~\eqref{Pearson_QI_eff} into the detection condition, and considering that the equality holds for the maximum detection range, we obtain
\begin{eqnarray}
	R_{\rm max} \times 10^{\frac{\gamma R_{\rm max}}{20}} = M^{\frac{1}{4}} \left[ \left(\frac{\tilde{\rho}_0^{\rm QI}}{\rho_{\mathrm{th}}} \right)^2 - 1 \right]^{\frac{1}{4}} R_c. \label{rangeEq_1_inequalityNoiseRdara}
\end{eqnarray}
By comparing this expression with its counterpart for {\color{black}the} direct detection radars obtained in Eq.~\eqref{max_range_equation}, we {\color{black}found} that the term
$\big[  (\tilde{\rho}_0^{\rm QI}/\rho_\mathrm{th})^2 - 1  \big]^{1/4}$
on the right-hand side of Eq.~\eqref{rangeEq_1_inequalityNoiseRdara} analogously {\color{black}plays the role of} the term
$(\mathrm{SNR}_{\mathrm{th}})^{-1/4}$
{\color{black}which is} appeared {\color{black}on} the right-hand side of Eq.~\eqref{max_range_equation} for direct detection radars. This analogy motivates {\color{black}us to define} an effective threshold SNR for {\color{black}the} target detection in quantum{\color{black}-entangled} noise radars as
\begin{eqnarray}
	\mathrm{SNR}_{\mathrm{th,eff}}^{\mathrm{QI}} \equiv \left[ \left( \frac{\tilde{\rho}_0^{\rm QI}}{\rho_{\mathrm{th}}} \right)^2 - 1 \right]^{-1}. \label{SNR_th_NR}
\end{eqnarray}
{\color{black}This definition allows us} {\color{black}to interpret} the quantum-entangled {\color{black}noise} radar as a direct detection radar with the effective threshold SNR that obtained by Eq.\eqref{SNR_th_NR}.

\begin{figure*}[!htb]
	\centering
	\begin{subfigure}[b]{0.47\textwidth}
		\includegraphics[width=\textwidth]{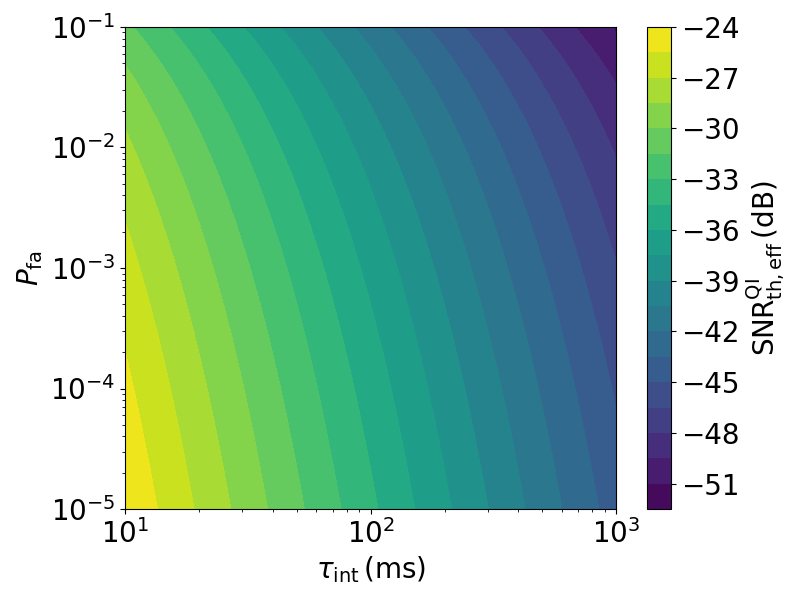}
		\caption{}
		\label{Fig6a}
	\end{subfigure}
	\hfil
	\begin{subfigure}[b]{0.47\textwidth}
		\includegraphics[width=\textwidth]{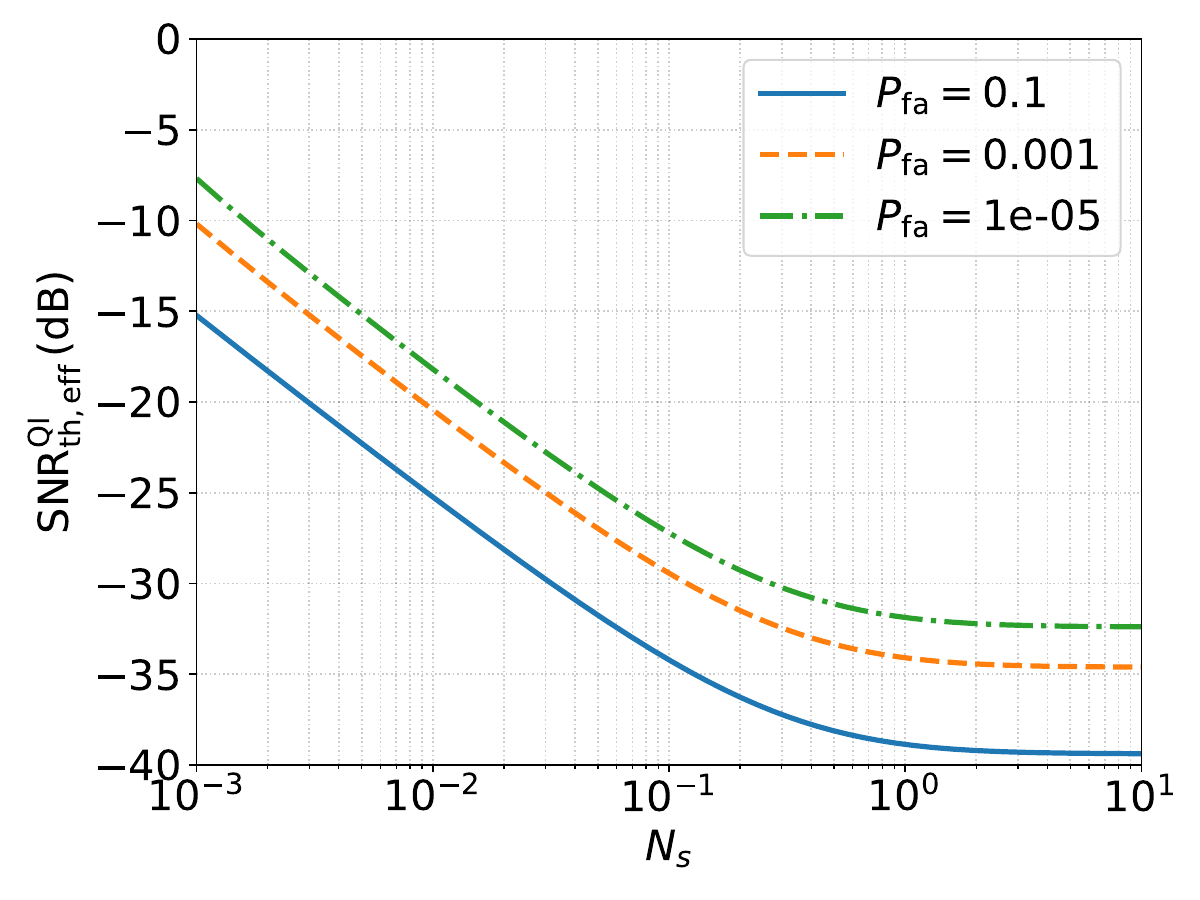} 
		\caption{}
		\label{Fig6b}
	\end{subfigure}
	\caption{ (Color online)  The effective threshold $\rm SNR$ for {\color{black}the} quantum entangled noise radars ${\rm SNR}_{\rm th, eff}^{\rm QI}$ in term of \textbf{(a)}  the false-alarm probability of {\color{black}the} target detection $P_{\rm fa}$ and the integration time $\tau_{\rm int}$ for {\color{black}fixed value} $N_s=0.1$, and  \textbf{(b)} the mean photon number per mode $N_s$ for $\tau_{\rm int}=100~\rm ms$ and different values of the false-alarm probability as $P_{\rm fa}=\{ 10^{-1}, 10^{-3}, 10^{-5} \}$. The {\color{black}other} system parameters considered here are {\color{black}the} same as {\color{black}in} Fig.~\ref{fig5}. }
	\label{Fig6}
\end{figure*}

{\color{black}In Fig.~\ref{Fig6}, we have shown} the influence of different parameters of the system, such as {\color{black}the} integration time $\tau_{\rm int}$, false-alarm probability $P_{\rm fa}$, and mean signal-idler photon number per mode $N_s$ {\color{black}on the} effective threshold SNR of {\color{black}quantum-entangled noise radars (Eq.~\eqref{SNR_th_NR})}.
As shown in {\color{black}Fig.~\ref{Fig6}}, the effective threshold SNR for target detection in quantum-entangled {\color{black}noise} radars is significantly lower than {\color{black}that of classical direct-detection} radars, which are typically greater than $\rm 10~dB$ \cite{skolnikBook}.
As discussed earlier in Sec.~\ref{sec2:direct_detection_radars}, the maximum detection range {\color{black}of a} direct-detection radar increases as the threshold SNR decreases. Since a quantum-entangled noise radar effectively behaves as a direct-detection radar with {\color{black}a reduced} threshold SNR, we expect its maximum detection range to {\color{black}exceed that of analogous} classical direct-detection radars with {\color{black}the} same system parameters.
Figure~\ref{Fig6a} demonstrates that $\mathrm{SNR}_{\mathrm{th,eff}}^{\mathrm{QI}}$ decreases with increasing $\tau_{\mathrm{int}}$ and $P_{\mathrm{fa}}$. Moreover, Figure~\ref{Fig6b} shows that increasing $N_s$ leads to a reduction {\color{black}of} the effective threshold SNR for target detection.

By substituting Eq.~\eqref{SNR_th_NR} into Eq.~\eqref{rangeEq_1_inequalityNoiseRdara}, and solving the resulting equation for $R_{\rm max}$, {\color{black}one can obtain} the maximum detection range of the quantum-entangled {\color{black}noise} radar in terms of the Lambert W function as {\color{black}follows}
\begin{eqnarray} 
	&  R_{\rm max}^{\rm QI} = \frac{20}{\rm ln(10) \gamma}   W_0\bigg[ \frac{\rm ln(10)}{ 20 } \gamma \bigg(\dfrac{M}{{{\rm SNR}_{\rm th}^{\rm QI}}} \bigg)^{1/4} R_c \bigg].   \label{rangeEqNR_final_2}
\end{eqnarray}
This equation {\color{black}allows us} {\color{black}to} evaluate {\color{black}the} {\color{black}performance} of a quantum-entangled 
noise radar by considering {\color{black}the} system, environment, and target {\color{black}parameters}. In the next subsection, we {\color{black}will} calculate the analogous relation for {\color{black}classical-correlated noise radars} {\color{black}to} compare their performance {\color{black}under} the same operational conditions.

\subsection{ Maximum Detection Range of Classical-Correlated Noise Radars} \label{Classical_noise_radar}

In an ideal classical{\color{black}-correlated} noise radar, {\color{black}the} correlated signal and idler fields {\color{black}are produced} {\color{black}by} splitting a coherent state of {\color{black}a} microwave field {\color{black}by means of} a power divider or beam splitter. 
{\color{black}{\color{black}A} power divider is characterized by its complex transmission $t$ and reflection $r$ coefficients (satisfying $|t|^2 + |r|^2 = 1$), and the relative phase {\color{black}$\varphi$ between} the transmission and reflection paths.}
{\color{black}The two input fields {\color{black}that incident on} the input ports of the power divider described by the annihilation operators $\hat{a}_1$ and $\hat{a}_2$. The output field operators of the power divider, i.e., the signal and idler ports, denoted by $\hat{a}_{\rm S}$ and $\hat{a}_{\rm I}$, {\color{black}respectively}, are related to the input operators via the following transformation} \cite{makarov2022theory}
\begin{eqnarray}
	\begin{pmatrix}
		\hat{a}_{\rm S} \\
		\hat{a}_{\rm I}
	\end{pmatrix}
	=
	\begin{pmatrix}
		|t| & e^{i\varphi} |r| \\
		- e^{-i\varphi} |r| & |t|
	\end{pmatrix}
	\begin{pmatrix}
		\hat{a}_{1} \\
		\hat{a}_{2}
	\end{pmatrix}.
\end{eqnarray}
{\color{black}This relation} describes the coherent mixing of the input modes and determines the quantum state of the output fields.

{\color{black}{\color{black}Suppose} that the quantum state of the input port 1 being the coherent state $\ket{\alpha}$ with $\alpha = |\alpha| e^{i\theta}$, and the input port 2 being the vacuum state $\ket{0}$. Thus, the joint quantum state of the input fields is given by $\ket{\Psi}_{\rm in}=\ket{\alpha}_1 \otimes \ket{0}_2$.}
Analogous to the quantum scheme, we should calculate the Pearson correlation coefficient {\color{black}that previously defined} in Eq.~\eqref{ro_1} to evaluate the correlation between these fields. By assuming an ideal balanced power divider, i.e. $ |t|^2 = |r|^2 =1/2$, and $\varphi=\pi$ ($\pi$ radians phase shift during reflection), {\color{black}considering the} phase of the coherent state as $\theta=0$ without the loss of generality, {\color{black}and also using the joint quantum state $\ket{\Psi}_{\rm in}$}, we get to
\begin{eqnarray} 
	& \braket{\hat{I}_{\rm S}^2}_{\rm CI} \simeq \eta(R) G_{\rm S} (|\alpha|^2+\frac{1}{2}) + N_{\rm S, add}^{(1)}, \\
	& \braket{\hat{I}_{\rm I}^2}_{\rm CI} \simeq G_{\rm I} (|\alpha|^2 + \frac{1}{2}) + N_{\rm I, add},\\
	& \braket{\hat{I}_{\rm S}\hat{I}_{\rm I}}_{\rm CI} = \sqrt{\eta(R) G_{\rm S} G_{\rm I}} |\alpha|^2.
\end{eqnarray}
By considering {\color{black}that} {\color{black}$|\alpha|^2=N_s$} {\color{black}in order} to have {\color{black}an} analogy between {\color{black}the} classical and quantum {\color{black}cases}, one  {\color{black}finds} $\braket{\hat{I}_{\rm S}^2}_{\rm CI}=\braket{\hat{I}_{\rm S}^2}_{\rm QI}$ and $\braket{\hat{I}_{\rm I}^2}_{\rm CI}=\braket{\hat{I}_{\rm I}^2}_{\rm QI}$, while $\braket{\hat{I}_{\rm S}\hat{I}_{\rm I}}_{\rm QI} =  Q_{\rm adv}\times \braket{\hat{I}_{\rm S}\hat{I}_{\rm I}}_{\rm QI}$ {\color{black}where}
\begin{align}  \label{Q_adv}
	& Q_{\rm adv}=\sqrt{1+\frac{1}{N_s}},
\end{align}
denotes the quantum advantage. By noticing Eq.~\eqref{ro_1} {\color{black}one} can conclude that the Pearson correlation coefficient for the quantum-entangled fields is $Q_{\rm adv}$ times greater than {\color{black}the classically correlated ones}. Therefore, {\color{black}the} Pearson correlation coefficient for the {\color{black}classically correlated {\color{black}fields} can be written as}
\begin{eqnarray}  
	& \rho_{\rm CI}^{\rm eff} (R) \simeq \frac{ \tilde{\rho}_0^{\rm CI} }{\sqrt{ 1+ \frac{1}{M G_{\rm S,rec} F^2(R)} \big(\frac{R}{R_c}\big)^4 } }, \label{Pearson_Classic}
\end{eqnarray}
{\color{black}where} $\tilde{\rho}_0^{\rm CI} = \tilde{\rho}_0^{\rm QI}/Q_{\rm adv}$ is the Pearson correlation coefficient of the classical correlated fields. 
{\color{black}Equation~\eqref{Pearson_Classic} shows that under {\color{black}the} identical conditions, the correlation between the two quantum-entangled signal-idler fields prepared in the TMSV state is enhanced by the factor of $Q_{\rm adv}$ compared to the ideal classical-correlated fields.}
{\color{black}It should be noted that the quantum advantage appeared here originates {\color{black}from the} quantum entanglement, which never has the classical counterpart.}

By applying the detection condition {\color{black}analogous to the quantum-entangled {\color{black}noise} radar}, {\color{black}we have derived} {\color{black}an explicit expression for the} maximum detection range of the {\color{black}classical-correlated} noise radar {\color{black}as follows}
\begin{eqnarray} 
	R_{\rm max}^{\rm CI} = \frac{20}{\ln(10) \gamma} W_0\left[  \frac{\ln(10)}{20}\gamma \left( \frac{M}{{\rm SNR}_{\rm th}^{\rm CI}} \right)^{1/4} R_c \right], \label{rangeEqNR_final_3}
\end{eqnarray}
where ${\rm SNR}_{\rm th}^{\rm CI}$ denotes the effective threshold SNR for the classical illumination target detection, and is given by
\begin{eqnarray} 
	{\rm SNR}_{\rm th}^{\rm CI} = \left[ \left( \frac{\tilde{\rho}_0^{\rm CI}}{\rho_{\mathrm{th}}} \right)^2 - 1 \right]^{-1}. \label{SNR_th_NR_2}
\end{eqnarray}
We are now {\color{black}going} to compare the maximum detection ranges of a {\color{black}classical-correlated} noise radar and {\color{black}a} quantum-entangled noise radar {\color{black}with} identical system parameters {\color{black}in the following subsection}.


\subsection{Range Enhancement Factor} \label{subsec3-3.Comparision}

%
%
%
\begin{figure*}[!htb]
	\centering
	\begin{subfigure}[b]{0.47\textwidth}
		\includegraphics[width=\textwidth]{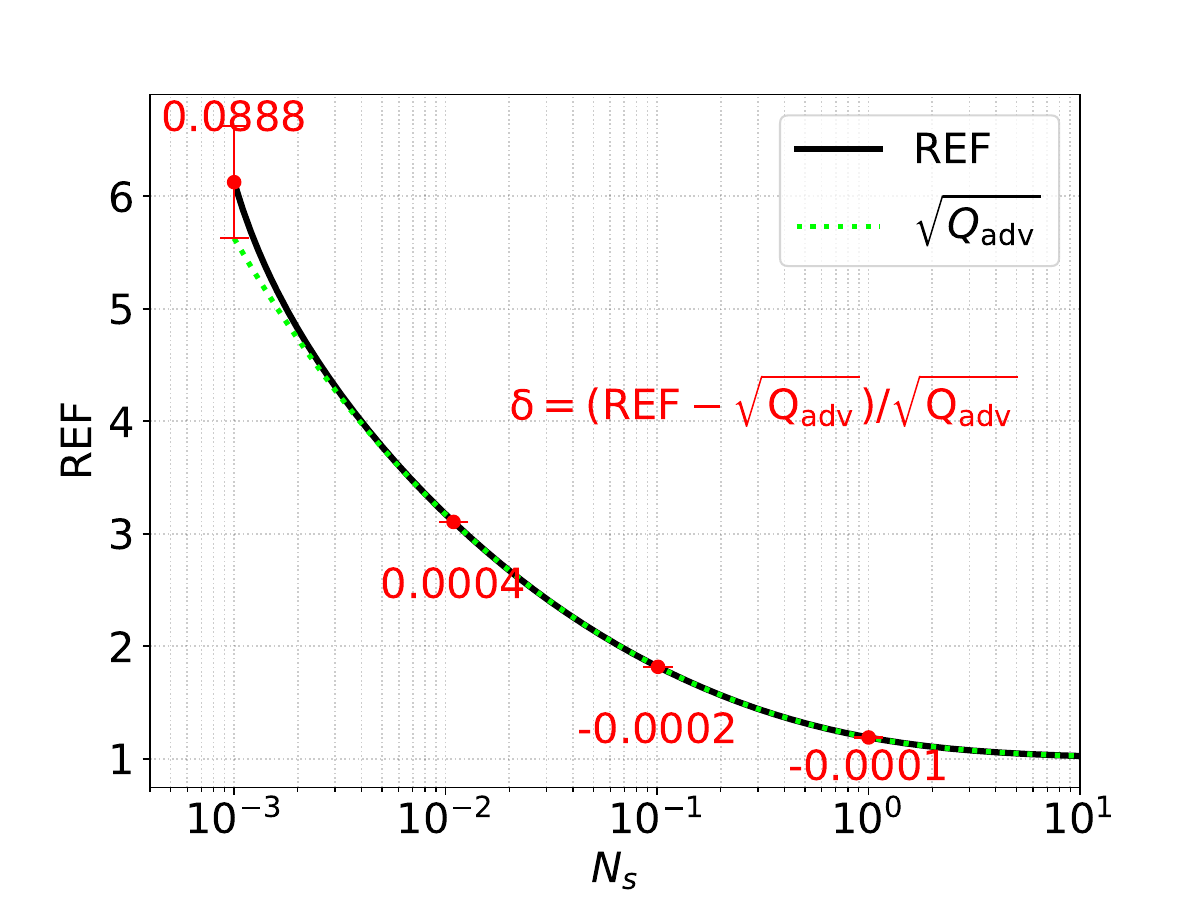}
		\caption{}
		\label{Fig7a}
	\end{subfigure}
	\begin{subfigure}[b]{0.47\textwidth}
		\includegraphics[width=\textwidth]{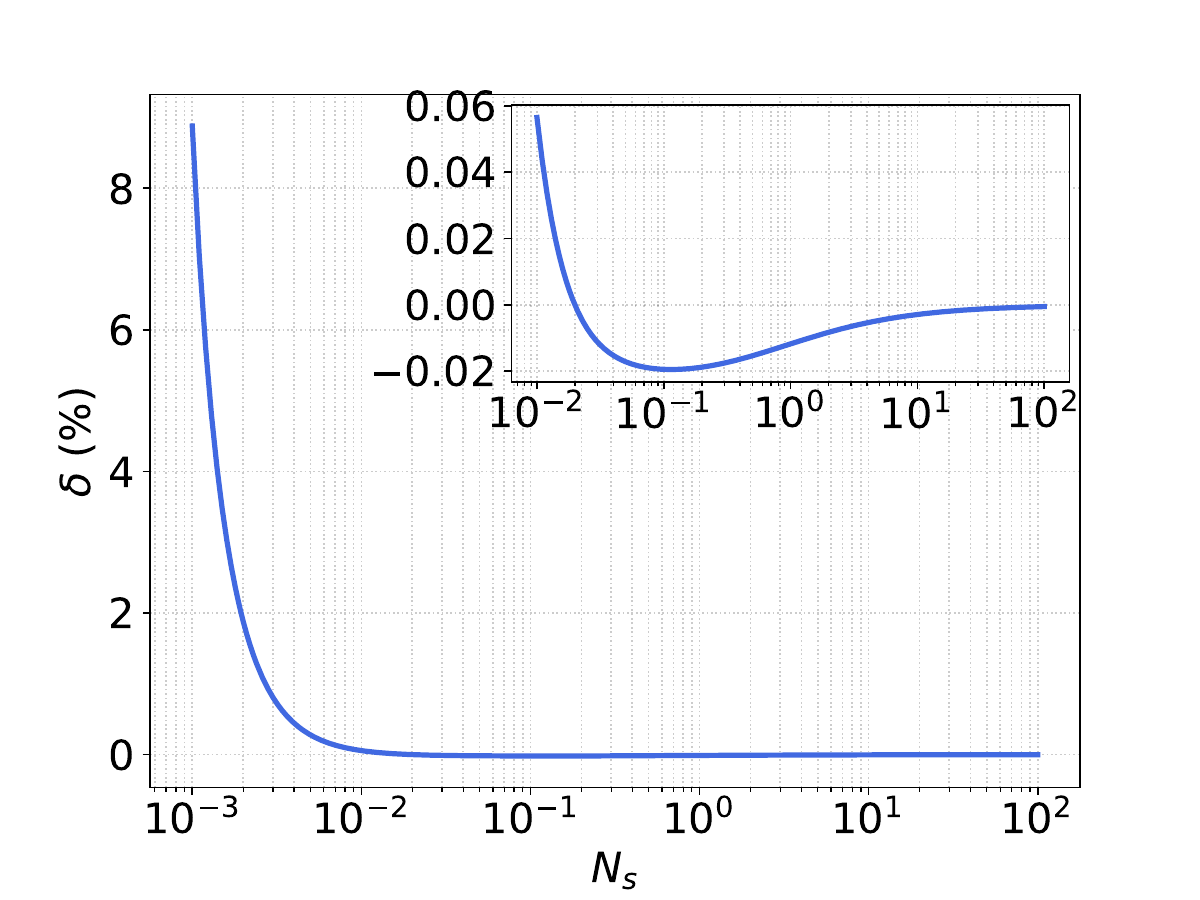} 
		\caption{}
		\label{Fig7b}
	\end{subfigure}
	\caption{(Color online) (a) {\color{black}The} range enhancement factor {\color{black}, $\rm REF$, versus the photon number per mode}, $N_s$, based on the analytical expression obtained in Eq.~\eqref{REF_1} (solid-black line) and {\color{black}the} approximation result given in Eq.~\eqref{REF_2} (dotted-green line). The difference $\delta$ {\color{black}is shown with the vertical red error bars for $N_s=$ $10^{-3}$, $10^{-2}$, $10^{-1}$, and $1$}. 
		(b) The difference between the exact analytical and approximate expressions {\color{black}for $\rm REF$}, i.e., $\delta$ {\color{black}vs} $N_s$. {\color{black}the other s}ystem parameters considered here are $f_{\rm S}=f_{\rm I}=9~\rm GHz$, $B_{\rm sig}=500~\rm MHz$.    }
	\label{Fig7}
\end{figure*}
%

{\color{black}In order to gain insight into the performance of quantum-entangled noise radars, it is beneficial to compare their maximum detection range with that of classical ones under identical operational conditions. To this end, we have defined} {\color{black}the parameter \textit{range enhancement factor} ($\rm REF$) as the ratio of the maximum detection range of a quantum-entangled noise radar to that of a classically correlated one, i.e., $\mathrm{REF}\equiv R_\mathrm{max}^\mathrm{QI}/ R_\mathrm{max}^\mathrm{CI}$,} {\color{black}for identical system parameters. From} {\color{black}Eqs.~\eqref{rangeEqNR_final_2} and \eqref{rangeEqNR_final_3}, the $\rm REF$ can be} {\color{black}obtained as follows}
\begin{eqnarray} 
	\mathrm{REF} = \frac{W_0\bigg[ \frac{\rm ln(10)}{ 20 } \gamma \bigg(\dfrac{M}{{{\rm SNR}_{\rm th}^{\rm QI}}} \bigg)^{1/4} R_c \bigg]}{W_0\bigg[ \frac{\rm ln(10)}{ 20 } \gamma \bigg(\dfrac{M}{{{\rm SNR}_{\rm th}^{\rm CI}}} \bigg)^{1/4} R_c \bigg]}. \label{REF_1}
\end{eqnarray}
However, since $W_0 (x)=\sum_{n=1}^{\infty}{ (-1)^{n-1}x^n/n! }$ \cite{steinvall2008laser}, for $\gamma\ll1$ {\color{black}one} can approximate the the Lambert W-function $W_0$ in Eq.~\eqref{REF_1} {\color{black}by} $W_0(x)\approx x$. Therefore, for $\gamma \ll 1$ {\color{black}one} can conclude that
\begin{eqnarray}\label{REF_SNR1}
	&&{\rm REF} \approx \bigg( \frac{ {\rm SNR}_{\rm th}^{\rm CI} }{{\rm SNR}_{\rm th}^{\rm QI}} \bigg)^{1/4}. 
\end{eqnarray}
On {\color{black}the} other hand, from Eqs.~\eqref{SNR_th_NR} and \eqref{SNR_th_NR_2}, {\color{black}one can show, after some straightforward calculations, that}
\begin{eqnarray} 
	\frac{\mathrm{{SNR}_{th}}^\mathrm{QI}}{\mathrm{{SNR}_{th}}^\mathrm{CI}} \propto 
	\begin{cases}
		1 & N_s \gg 1\\
		1/Q_\mathrm{adv}^{2} & N_s \ll 1
	\end{cases}. \label{SNR_th_ratio}
	\end{eqnarray}
{\color{black}Therefore,} {\color{black}in the limit of $\gamma \ll 1$, the range enhancement factor can be} {\color{black} approximated by} 
\begin{eqnarray} 
	\mathrm{REF} &\approx \sqrt{Q_{\rm adv}}. \label{REF_2}
\end{eqnarray}
{\color{black}This} relation provides a straightforward tool for {\color{black}comparing the maximum detection range of quantum-entangled noise radars with their classical counterparts.}

The difference between the exact analytical expression for $\mathrm{REF}$ obtained in Eq.~\eqref{REF_1} and its approximate value in Eq.~\eqref{REF_2} {\color{black}is} quantified by
\begin{eqnarray} 
	\delta = \frac{\mathrm{REF}-\sqrt{Q_{\rm adv}} }{\sqrt{Q_{\rm adv}} }. \label{REF_delta}
\end{eqnarray}

{\color{black}In order to examine} the validity of {\color{black}the approximations that were introduced previously}, we have plotted the exact analytical value of the $\mathrm{REF}$ obtained {\color{black}in Eq.~\eqref{REF_1}} and its approximate value in Eq.~\eqref{REF_2}, with respect to $N_s$ in Fig.~\ref{Fig7a}. This plot shows that there is good agreement between {\color{black}the exact value of} {\color{black}the} $\mathrm{REF}$ and its approximation {\color{black}obtained in Eq.~\eqref{REF_2}}. In this figure, the difference $\delta$ is displayed for some values of {\color{black}the} mean photon number per mode, {\color{black}as} $N_s=10^{-3}, 10^{-2}, 10^{-1}, 1$. {\color{black}It is evident that} the difference is just $0.0888$ for $N_s=10^{-3}$, and {\color{black}it is reduced} to less than $0.0004$ for larger values of $N_s$. Figure~\ref{Fig7b} {\color{black}gives} more insight into the variation of $\delta$ in terms of $N_s$. {\color{black}It} shows that the difference between {\color{black}the exact and approximate} {\color{black}values} of {\color{black}the} $\mathrm{REF}$ is about $8.9\%$ {\color{black}for} $N_s = 10^{-3}$, and it gradually vanishes as $N_s$ increases. {\color{black}This} {\color{black}demonstrates} that Eq.~\eqref{REF_2} gives a good approximation for the {\color{black}$\mathrm{REF}$}.

Let us now discuss the REF from the engineering point of view. As demonstrated previously in Eq.~\eqref{REF_SNR1}, for $\gamma \ll 1$, {\color{black}one can approximate} the REF as {\color{black}the fourth root of} the ratio between the threshold SNR of a {\color{black}classical-correlated} noise radar and that of a quantum-entangled one. It is straightforward to rewrite this equation as {\color{black}follows}
\begin{eqnarray}
	&&{\rm REF} \approx 2^{ \Delta {\rm SNR}_{\rm th}^{\rm (dB)}/12 } \label{REF_SNR2},
\end{eqnarray}
{\color{black}where $\Delta {\rm SNR}_{\rm th}^{\rm (dB)} \equiv {\rm SNR}_{\rm th}^{\rm CI (dB)} - {\rm SNR}_{\rm th}^{\rm QI (dB)}$}. {\color{black}According to} {\color{black}this equation, which makes practical sense, each 12dB reduction in the threshold SNR by means of the quantum entanglement {\color{black}doubles} the REF}.
%
From Eqs.~\eqref{REF_2}, \eqref{REF_SNR2}, and \eqref{Q_adv}, it is straightforward to show that for $N_s \ll 1$ (which is the case of quantum illumination), the parameter $\Delta {\rm SNR}_{\rm th}^{\rm (dB)}$ can {\color{black}be} approximated {\color{black}as}
\begin{eqnarray}
	&&\Delta {\rm SNR}_{\rm th}^{\rm (dB)} \approx  -3 \log_2(N_s) , \label{delta_SNR_th}
\end{eqnarray}

{\color{black}In conclusion}, Eqs~\eqref{REF_1}, \eqref{REF_2}, \eqref{REF_SNR2}, and \eqref{delta_SNR_th} provide a practical {\color{black}tool} from the engineering point of view to compare the performance of {\color{black}the quantum and classical versions}of noise radars.
{\color{black}As an example, for $N_s \simeq 0.25$ , we have $\Delta {\rm SNR}_{\rm th}^{\rm (dB)}= 6~\rm dB$}, which results in ${\rm REF} \simeq 1.41$, {\color{black}i.e., about $40\%$ superiority of the maximum detection range of the quantum-entangled noise radar over the classical-correlated one with the same operational conditions.}


\subsection{Results} \label{subsec3-4.results}
{\color{black}Previously}, we developed {\color{black}the} analytical framework for evaluating the maximum detection range of quantum-entangled noise radars, as well as their classical counterparts. 
{\color{black}This framework allows us to investigate} whether a quantum-entangled noise radar is capable of detecting targets at long ranges {\color{black}in {\color{black}the} order} of several kilometers. More specifically, {\color{black}it allows us to examine} the conditions {\color{black}under {\color{black}which}} such detection might be achievable.
{\color{black}To address this issue}, we assessed the maximum detection range of these systems using a set of realistic parameters {\color{black}based on} current technological capabilities, as summarized in Table~\ref{tab1}. It is important to note that we do not claim any experimental realization of {\color{black}a} quantum-entangled radar system operating with the parameters outlined in Table~\ref{tab1}.

\begin{table}[!htb]
	\caption{{\color{black}The} system parameters considered for a quantum-entangled noise radar.}
	\begin{center}
		\begin{tabular}{ccc}
			\toprule
			Parameter& Value & Unit \\
			\toprule
			Microwave Photon Pair Source  &  JTWPA & -\\
			$f_{\rm S}$  & 9 & GHz \\ 
			$f_{\rm I}$ &  9 & GHz   \\
			$B_{\rm sig}$  & 8 & GHz \\ 
			$G_{\mathrm{tr}}$  &  80 & dB \\
			$G_{\mathrm{S, rec}}$ &  20 & dB  \\
			$T_{\rm amp}$      & 10 & mK	\\
			NF	&	1	& $\rm dB$ \\
			$T_{\rm env}$ & 300 & K  \\ 
			$\tau_{\mathrm{int}}$ & 200 & ms  \\
			$B_{\mathrm{det}}$ & 50 & MHz \\ 
			$N_s$ & 0.05 & - \\
			$\varepsilon_a$ & 0.9 & -\\
			$A$ & $\pi$ & $\rm m^2$ \\
			$P_{\rm fa}$ (search mode)	& $10^{-1}$ & - \\
			$P_{\rm fa}$ (track mode)	& $10^{-3}$ & - \\
			\toprule
		\end{tabular}
	\end{center}
	\label{tab1}
\end{table}

\begin{figure*}[!htb]
	\centering
	\begin{subfigure}[b]{0.47\textwidth}
		\includegraphics[width=\textwidth]{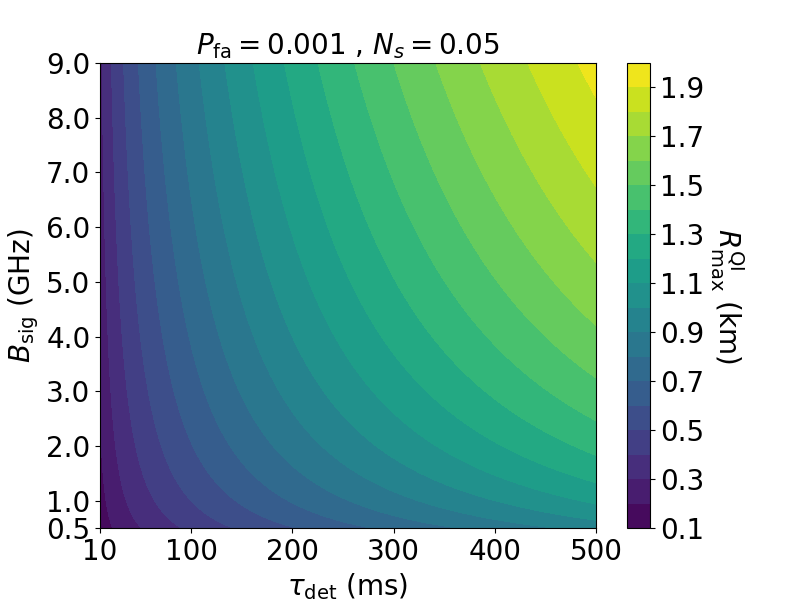}
		\caption{}
		\label{Fig8a}
	\end{subfigure}
	\hfill
	\begin{subfigure}[b]{0.47\textwidth}
		\includegraphics[width=\textwidth]{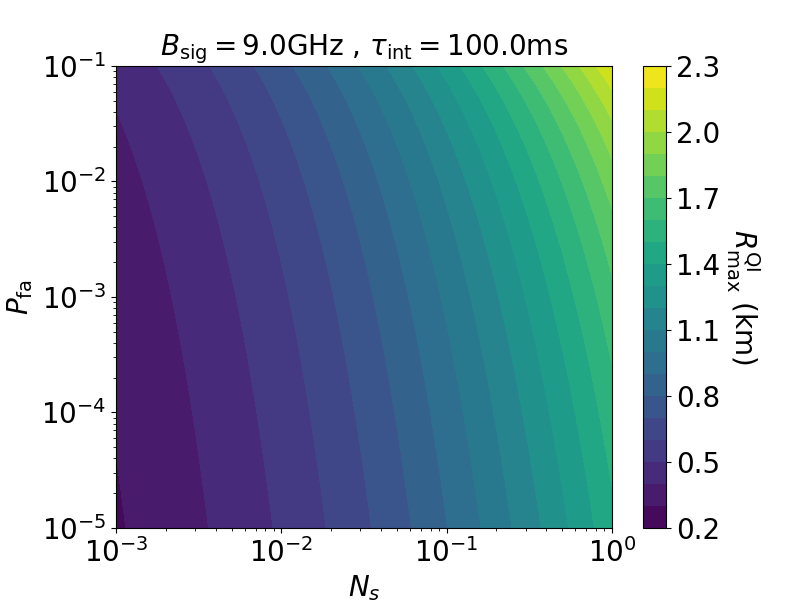} 
		\caption{}
		\label{Fig8b}
	\end{subfigure}
	\caption{(Color online) {\color{black}The} maximum detection range of {\color{black}the} quantum-entangled noise radar, $R_{\rm max}^{\rm QI}$, in terms of: \textbf{(a)} {\color{black}the} detection time $\tau_{ \mathrm{int}}$ and {\color{black}the} signal bandwidth $B_{\rm sig}$ for $P_{\rm fa}=10^{-3}$ and $N_s= 0.05$ ({\color{black}corresponding} to $\mathrm{REF}=2.14$), and \textbf{(b)} {\color{black}the} mean photon number per mode $N_s$ and {\color{black}the} false-alarm probability $P_{\rm fa}$ for $B_{\rm sig}=9~\rm GHz$ and $\tau_{ \mathrm{int}}=50~\rm ms$. {\color{black}The other system} parameters {\color{black}are the same as} given in Table.~\ref{tab1}. 
	} 
	\label{Fig8}
\end{figure*}

Following the approach outlined in {\color{black}Ref.}~\cite{liveri202210GHz, zorin2019flux}, we consider a quantum-entangled noise radar utilizing {\color{black}a} state-of-the-art flux-driven Josephson traveling-wave parametric amplifier (JTWPA) as the source of entangled microwave-photon pairs. This device is capable of emitting degenerate signal and idler fields at frequencies $ f_{\rm S} = f_{\rm I} = 9~\mathrm{GHz} $, with an average photon number per mode of $ N_s = 0.05 $, and a 3-dB bandwidth {\color{black}at} the signal center frequency.  
This performance arises from {\color{black}the} efficient three-wave mixing and broadband phase matching, which are enabled by the circuit’s nonlinear dispersion and the separate pump-signal configuration \cite{zorin2019flux}. Consequently, the signal bandwidth of $B_{\rm sig} = 8~\mathrm{GHz}$ considered in Table~\ref{tab1} is experimentally achievable with this technology. 
The total transmitter gain is assumed to be $ 80~\mathrm{dB}$, {\color{black}which is} consistent with {\color{black}that} reported in Ref.~\cite{barzanjehExperimentQuantumRadar2020}. Additionally, {\color{black}the} signal amplification gain at the receiver is considered {\color{black}to be} \( G_{\rm S, rec} = 20~\mathrm{dB} \).  
We assumed {\color{black}the} detection time as $\rm 200~ms$, which is appropriate for detecting medium-sized objects at velocities of approximately $\rm 20-30~m/s$ over distances up to $\rm 1~km$ \cite{mandula2024towards}. This is justified by the fact that the object's displacement during the detection interval is less than $0.60\%$ of its range, ensuring minimal {\color{black}impact of target movement} on {\color{black}the} detection.  

Two operational modes based on the target detection false-alarm probability are considered: track mode ($P_{\rm fa} < 0.01$) and search mode ($0.01 < P_{\rm fa} < 0.5$).  
According to Eq.~\eqref{rangeEqNR_final_2}, the quantum-entangled noise radar with the parameters listed in Table~\ref{tab1} is capable of detecting an object with an RCS of $\sigma = 0.1~\mathrm{m}^2$ at a maximum range of $R_{\rm max} = 1.56~\mathrm{km}$ in the search mode ($P_{\rm fa} = 0.1$) and $R_{\rm max} = 1.19~\mathrm{km}$ in the track mode ($P_{\rm fa} = 0.001$). 

{\color{black}It is important to emphasize that, for $N_s = 0.05$, the considered system shows ${\rm REF}\simeq 2.14$, $Q_{\rm adv}\simeq 4.58$, and $\rm \Delta {\rm SNR}_{th}^{\rm (dB)}= 13.18~\rm dB$. This refers to the fact that, by utilizing entanglement, the maximum detection range of the considered system exceeds by {\color{black}a} factor of $2.14$, and the threshold SNR is reduced by $13.18~\rm dB$, compare to {\color{black}its analogous classical-}correlated noise radar under the same operational conditions. Moreover, it shows that the signal and idler modes in the considered quantum-entangled noise radar {\color{black}have a correlation which is} $4.54$ times stronger than {\color{black}the classical case}.}

To provide further insight, we analyze the impact of the key system parameters on the maximum detection range of the considered quantum-entangled noise radar in Fig.~\ref{Fig8}. 
{\color{black}Figure~\ref{Fig8a} shows} the influence of the integration time $\tau_{ \mathrm{int}}$ and {\color{black}the} signal bandwidth $B_{\rm sig}$ on the maximum detection range of the quantum-entangled noise radar with parameters listed in Table~\ref{tab1} in the track mode with $P_{\rm fa}=0.001$. {\color{black}It also demonstrates} that with an ultra-wideband signal with $B_{\rm sig}\sim 9~\rm GHz$ and a sufficiently long detection time as $\tau_{ \mathrm{int}}=500~\rm ms$, tracking low reflective objects with $\sigma=0.1~\rm m^2$ {\color{black}up to} $2~\rm km$ is possible with the considered system. 
{\color{black}Moreover, it} shows the importance of the signal bandwidth on the maximum detection range of the quantum-entangled {\color{black}noise} radar. {\color{black}As an example}, for {\color{black}the} given integration time as $\tau_{\rm int}=100~\rm ms$, {\color{black}the} maximum detection range increases from $420~\rm m$ to $866~\rm m$ when the signal bandwidth rises from $500~\rm MHz$ to $9~\rm GHz$.
The effect of the false-alarm probability, $P_{\rm fa}$, and the mean photon number of the generated signal and idler fields per mode, $N_s$, on the maximum detection range is demonstrated in Fig.~\ref{Fig8b}. In this figure, we {\color{black}have} considered a signal with ultra-high bandwidth of $B_{\rm sig}=9~\rm GHz$ and a relatively short integration time as $\tau_{ \mathrm{int}}=100~\rm ms$. {\color{black}As is seen}, for $N_s=1$ ({\color{black}corresponding to} $Q_{\rm adv}=1.41$ and $\mathrm{REF}\simeq1.18$), the maximum detection range reaches $\rm 2.2~\rm km$ and $\rm 1.67~\rm km$ {\color{black}in the} search and track modes, respectively. 
However, for a low value of the photon {\color{black}number} per mode {\color{black}as} $N_s =0.01$ where the quantum advantage and $\mathrm{REF}$ {\color{black}are} substantial, the maximum detection range reaches {\color{black}to} $769~\rm m$ in the search mode and $584~\rm m$ in the track mode.

It is important to {\color{black}note} that the parameters proposed in Table~\ref{tab1} are practically achievable with {\color{black}the} current state-of-the-art {\color{black}technology}. To support this claim, we compare these parameters with those of the previously implemented quantum-entangled noise radars, as summarized in Table~\ref{tab3}. This table presents the key system parameters of three notable quantum radar {\color{black}implementations which are} reported to date\cite{barzanjehExperimentQuantumRadar2020, balaji2019receiver, livreri2023JTWPA}, to the best of our knowledge.
{\color{black}In order to} estimate the maximum detection range of these systems, we consider a transmission antenna characterized by {\color{black}the} gain of  $G = 15\,\mathrm{dB}$, {\color{black}the} aperture efficiency of $\varepsilon_a = 0.9$, and {\color{black}the} radius $r = 1\,\mathrm{m}$. The target is assumed to be a small object with {\color{black}the} RCS of $\sigma = 0.1\,\mathrm{m}^2$. {\color{black}The} false-alarm probability is set to $P_{\mathrm{fa}} = 0.1$ for the search mode and $P_{\mathrm{fa}} = 0.001$ for the track mode. {\color{black}The} atmospheric absorption coefficient of $\gamma = 0.007\,\mathrm{dB/km}$ {\color{black}is also considered}.
Under these conditions, the maximum detection range of the quantum-entangled noise radar reported in \cite{balaji2019receiver} is estimated to be $17.82\,\mathrm{m}$ in the search mode. For the system described in \cite{barzanjehExperimentQuantumRadar2020}, the estimated range increases to $378.6\,\mathrm{m}$, while the radar reported in \cite{livreri2023JTWPA} achieves {\color{black}the} maximum range of $354.42\,\mathrm{m}$. In the track mode, the corresponding maximum detection ranges are estimated as $13.54\,\mathrm{m}$ for \cite{balaji2019receiver}, $285.75\,\mathrm{m}$ for \cite{barzanjehExperimentQuantumRadar2020}, and $269\,\mathrm{m}$ for \cite{livreri2023JTWPA}.

As shown in Table~\ref{tab3}, the system presented in {\color{black}Ref.~\cite{livreri2023JTWPA} has} the highest quantum advantage, $Q_{\mathrm{adv}}$, among the three. This enhanced performance is primarily attributed to the lower signal photon number $N_s$ utilized in the system. While the radar demonstrated in \cite{balaji2019receiver} yields the lowest performance in terms of both maximum detection range and quantum advantage, it offers a notably short integration time, making it more aligned with real-time field applications.
Furthermore, the {\color{black}longer} detection range of the system in Ref.~\cite{livreri2023JTWPA} is {\color{black}due to} its high signal bandwidth. {\color{black}In addition, it} could be further improved through the use of a higher-gain {\color{black}amplifier} at the transmitter, {\color{black}as well as by increasing the transmitted signal photon number per mode} $N_s$.

{\color{black}It is worthwhile to {\color{black}note} that} the objective of this study is to evaluate the ultimate performance of quantum-entangled noise radars. For this purpose, {\color{black}up to now we have assumed} that the receiver antenna is ideal, capable of detecting any weak signal incident upon it. In reality, however, various microwave {\color{black}detection} technologies have limitations in detecting weak signals, which may {\color{black}have an} impact {\color{black}on} the performance of the radar system. This issue will be addressed in the subsequent subsection \ref{subsec3.5-challenges}.

\begin{table}
	\centering
	\caption{List of the system parameters for some representative quantum-entangled radars implemented to-date (based to our knowledge). JPC: Josephson Parametric Converter; JPA: Josephson Parametric Amplifier; JTWPA: Josephson Traveling Wave Parametric Amplifier.}
	\begin{tabular}{ccccc}
		\toprule
		Parameter&  
		Unit & 
		Ref. \cite{barzanjehExperimentQuantumRadar2020} &
		Ref. \cite{balaji2019receiver} &
		Ref. \cite{livreri2023JTWPA} \\
		\toprule
		Source  & - & JPC & JPA & JTWPA\\
		$f_{\rm S}$ & GHz & 10.09 & 7.5376 & 3.3   \\ 
		$f_{\rm I}$ & GHz & 6.8 & 6.1445 & 3.45  \\
		$B_{\rm sig}$ & MHz & 20 & 1 & 3,000  \\ 
		$G_{\mathrm{tr}}$ & dB & 77.16 & 63 & 30  \\
		$N_b$ & - & 672 & 1015 & 100  \\ 
		$\tau_{\mathrm{int}}$ & ms & 1,900 & 50 & -  \\
		$B_{\mathrm{det}}$ & kHz & 200 & 1,000 & -  \\ 
		$M$ & - & $3.8\times10^5$ & $5\times10^4$ &  $1.5\times10^7$ \\
		$N_s$ & - & 0.5 & 0.57 & 0.05 \\
		$Q_{\rm adv}$ & - & 1.73 & 1.65 & 4.58 \\
		$\mathrm{REF}$ & - & 1.32 & 1.29 & 2.14 \\
		\toprule
	\end{tabular}
	\label{tab3}
\end{table}

\subsection{Challenges and Limitations} \label{subsec3.5-challenges} 
Thus far, we have comprehensively {\color{black}examined} the practical aspects of quantum-entangled noise radars. We have also investigated the feasibility of implementing such systems with a maximum detection range {\color{black}in} the order of few kilometers. However, several operational challenges remain, as identified by the quantum radar community, which we {\color{black}are going to address} in the following.

\subsubsection{Power of the Received Signal} \label{subsubsec3.5.1_received power}

One important concern regarding {\color{black}the} {\color{black}operation} of quantum-entangled noise radars is whether the power of the signal {\color{black}back-scattered} from the target and reaches the radar receiver is sufficient to be detected.
The radar receiver antenna can detect any signal whose power exceeds a fundamental physical detection limit (PDL). {\color{black}It is} defined as the minimum incident electromagnetic power required to induce a measurable electrical current or voltage at the antenna terminals. This limit is fundamentally determined by the antenna’s material properties, geometry, and quantum effects, and is independent {\color{black}of} the receiver system or signal processing techniques. 

For microwave antennas, the PDL corresponds to the power level at which {\color{black}the} conduction current induced by the incident wave becomes comparable to the intrinsic thermal agitation of electrons within the antenna structure. This establishes a bound below which no physical signal can be observed~\cite{skolnikBook}.

It is important to distinguish the PDL from both the antenna noise floor and the MDS defined in Eq.~\eqref{MDS_formula}. The noise floor refers to the total noise power present at the receiver input, which includes thermal noise contributions from the antenna and front-end electronics. {\color{black}Therefore, it defines} the practical sensitivity threshold for signal detection. MDS is a higher threshold that incorporates the system noise figure and the required {\color{black}SNR} for achieving {\color{black}the} specified detection and false alarm probabilities. In contrast to the noise floor and MDS, the PDL is an intrinsic property of the antenna and cannot be reduced through improvements in {\color{black}the} receiver design or {\color{black}the} signal processing.

It should be emphasized that in certain radar systems, such as noise radars (either classical or quantum) or those employing coherent processing and pulse coded or shaped waveforms, thanks to the processing gain, signals below the noise floor can be detected using advanced techniques such as integration, correlation, or matched filtering. 
However, such processing gains are ultimately constrained by the PDL of the antenna or detector. If the incident signal power is below this fundamental threshold, then it fails to induce any measurable electrical response. Therefore, it is trivial that no signal processing technique can recover it, as the signal does not physically exist in the received output.

Unlike the noise floor {\color{black}and} MDS, the {\color{black}PDL} of an antenna is not typically expressed by a simple universal formula. However, an approximate physical interpretation can be obtained by modeling the antenna as a resistor at temperature $T_a$, characterized by a Johnson-Nyquist noise current {\color{black}which is} given by \cite{johnson1928thermal}
\begin{align}
	i_n = \sqrt{\frac{4 k_B T_a B_a}{R_a}}. \label{Eq. Johnson-Nyquist noise current}
\end{align}
Here, $i_n$ is the root mean square (RMS) noise current, $B_a$ is the antenna bandwidth, and $R_a$ is the antenna radiation resistance. {\color{black}If a signal with the power $P$ incidents on an antenna, then the induced signal current is determined by}
\begin{equation}
	i_s = \sqrt{\frac{P}{R_a}}. \label{Eq. signal_current}
\end{equation}
{\color{black}The PDL corresponds to the minimum signal power for which the induced signal current, $i_s$, exceeds the noise {\color{black}current, $i_n$}. Therefore, by utilizing Eqs.~\eqref{Eq. Johnson-Nyquist noise current} and \eqref{Eq. signal_current}, one can obtain the PDL of classical antennas as {\color{black}follows}}
\begin{equation}
	P_\mathrm{PDL}^{\rm CA} \approx 4 k_B T_a B_a. \label{Eq. PDL}
\end{equation}
For instance, a high-performance antenna with an effective noise temperature of $T_a = 30~\mathrm{K}$ and a bandwidth of $B_a = 50~\mathrm{MHz}$ exhibits a PDL of $P_{\mathrm{PDL}}^{\rm CA} = 8.3 \times 10^{-14}~\mathrm{W}$ ($-100.8~\rm dBm$). In contrast, a commercial antenna operating at $T_a = 290~\mathrm{K}$ with the same bandwidth has a PDL of $P_{\mathrm{PDL}}^{\rm CA} = 8.0 \times 10^{-13}~\mathrm{W}$ ($-90.97~\rm dBm$).

In the case of SMPDs, the PDL can be estimated based on the detector's DCR. {\color{black}In this case}, the detector's PDL is limited by the DCR-equivalent power. By considering the quantum efficiency of the SMPD as $\eta_Q^{\rm SMPD}$, the PDL can be approximated {\color{black}as}
\begin{equation}
	P_\mathrm{PDL}^{\rm \rm SMPD} \approx \frac{1}{\eta_Q^{\rm SMPD}} hf_{\rm S} \times\mathrm{DCR}_{\rm SMPD}, \label{Eq. PDL_SMPD}
\end{equation}
{\color{black}where $f_{\mathrm{S}}$ is} the signal frequency. {\color{black}According to this relation}, the PDL for the SMPD reported in {\color{black}Ref.}~\cite{pallegoix2025enhancing} with {\color{black}a DCR of} $30~\mathrm{Hz}$ and the quantum efficiency of $\eta_Q^{\rm SMPD}=0.8$ {\color{black}can be found as} $P_{\mathrm{PDL}}^{\rm SMPD} \approx 2.12 \times 10^{-22}~\mathrm{W}$ ($-186.55~\rm dBm$).

\begin{figure*}[!htb]
	\centering
	\begin{subfigure}[b]{0.47\textwidth}
		\includegraphics[width=\textwidth]{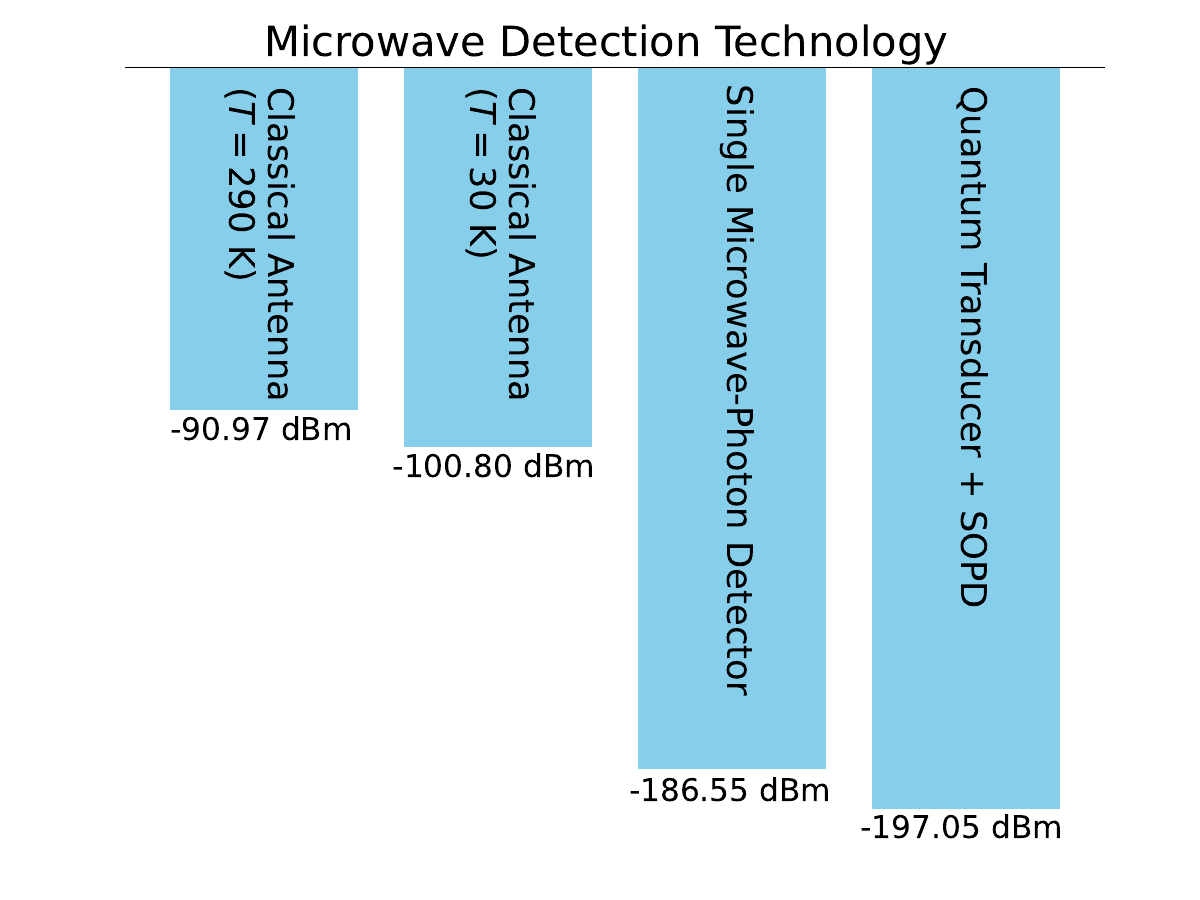}
		\caption{}
		\label{Fig9a}
	\end{subfigure}
	\hfill
	\begin{subfigure}[b]{0.47\textwidth}
		\includegraphics[width=\textwidth]{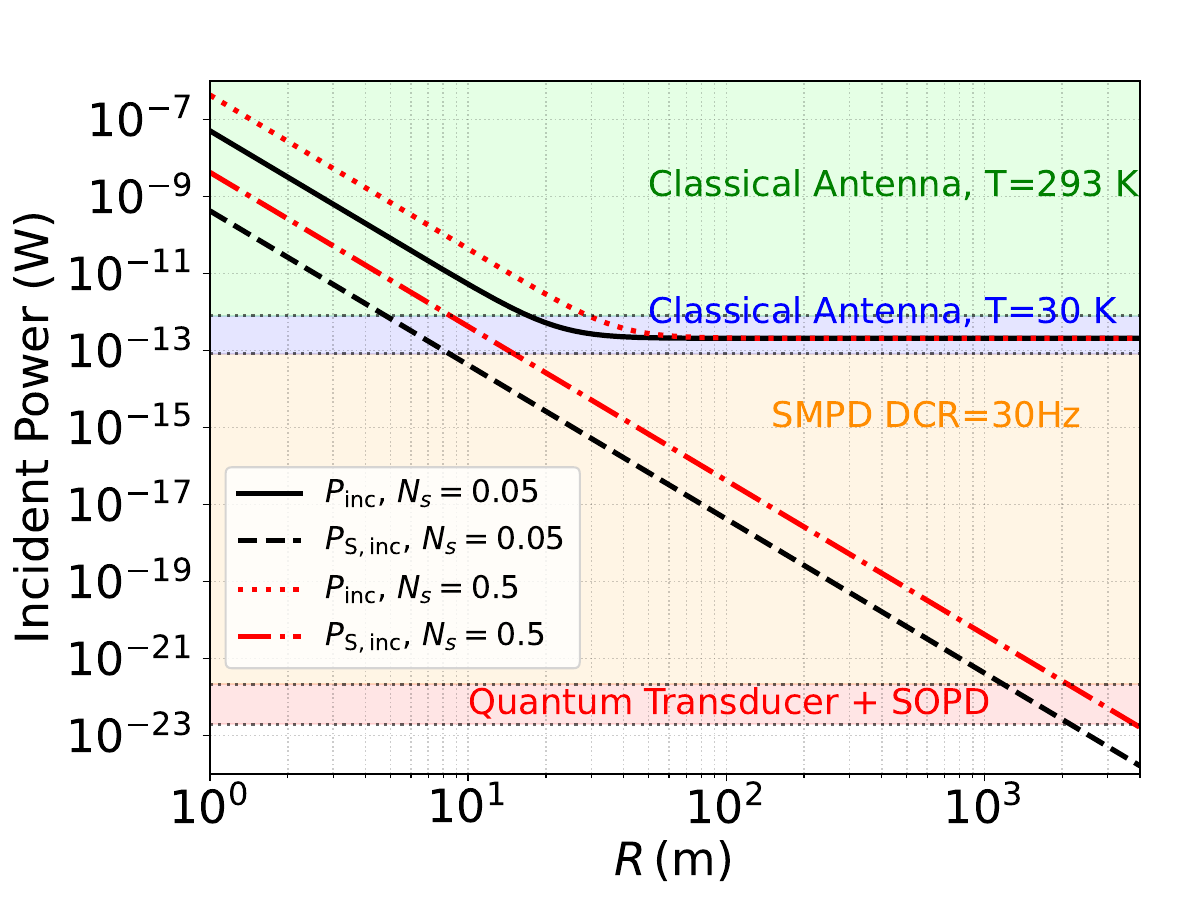} 
		\caption{}
		\label{Fig9b}
	\end{subfigure}
	\caption{ (Color online) (a) Comparison of the physical detection limit of various microwave detection technologies. (b) {\color{black}The} power of the total microwave field (signal + noise) and the signal field incident on the radar receiver as {\color{black}the} function of target range $R$ for {\color{black}the} different values of $N_s$ as 0.05 and 0.5. The system parameters used in this figure are {\color{black}the same as} in Table~\ref{tab1}. The shaded regions indicate the operational regimes of {\color{black}the} different microwave detection technologies.
	}
	\label{Fig9}
\end{figure*}

Another solution for detecting extremely weak microwave signals efficiently and  preserving its quantum state simultaneously, is based on quantum microwave-to-optical transducers, as {\color{black}denoted in Ref.}~\cite{lauk2020perspectives}.
{\color{black}In this method, the weak microwave signal arrives at the radar receiver should be coherently converted to the optical domain using a quantum microwave-to-optical transducer, and subsequently, the optical signal is detected by utilizing a high quantum-efficiency SOPD.}
Quantum transducers with high conversion efficiency of approximately $47\%$ and very low added noise below one photon based on  silicon nanomechanics technology have been reported in Ref.\cite{zhao2025quantum}. On the other hand, the technology of SOPDs is now matured, and SOPDs with detection efficiency of up to $90\%$ and low DCR as low as $1~\rm Hz$ are now available commercially (for instance, see IDQ ID281 SNSPD). 
{\color{black}Suppose} the PDL of the SOPD as its DCR-equivalent power, and consider {\color{black}that} the conversion efficiency of the coherent quantum microwave-to-{\color{black}optical} transducer is denoted by $\eta_{\rm trans.}$. {\color{black}Then}, the PDL for the efficient microwave detection scenario based on the quantum transducer is given by
\begin{equation}
	P_\mathrm{PDL}^{\rm trans.} \approx \frac{1}{\eta_Q^{\rm SOPD}\eta_{\rm trans.} } hf_{\rm S} \mathrm{DCR}_{\rm SOPD}. \label{Eq. PDL_trans}
\end{equation}
{\color{black}As an example, by employing}  the SOPD with $\eta_Q^{\rm SOPD} = 0.9$ and $\mathrm{DCR}_{\rm SOPD}=1~\rm Hz$, and the quantum microwave-to-optical transducer with {\color{black}the} conversion efficiency of $\eta_{\rm trans.} = 0.47$, one can obtain the PDL using Eq.~\eqref{Eq. PDL_trans} as $P_\mathrm{PDL}^{\rm trans.}\approx 1.97\times10^{-23}~\rm W$ ($-197.05~\rm dBm$). 
Figure~\ref{Fig9a} provides a comparative illustration of the PDL of different microwave detection technologies for better intuition. 

{\color{black}Based on the aforementioned analyses, we are now going to evaluate the performance of the quantum-entangled noise radars by considering the restrictions induced by the microwave detection technologies. This allows us to determine the most efficient microwave detection scenario that preserves {\color{black}the} advantages {\color{black}provided by} quantum-entangled noise radars.
To address this, we have to compare the power of the received signal with the PDL of the various microwave detection technologies introduced in this section.}

{\color{black}It is obvious that the} total number of photons {\color{black}that} incident on the receiver antenna is given by
\begin{equation}
	N_{\rm inc} = \braket{ \hat{a}_{\rm S, inc}^\dagger \hat{a}_{\rm S, inc} }, \label{Eq. Received number}
\end{equation}
where $\hat{a}_{\rm S, inc}$ denotes the annihilation operator of the incident microwave field and is given as
\begin{align}
	\hat{a}_{\rm S, inc} ={}& \sqrt{\eta(R) G_{\rm S, tr}}\, \hat{a}_{\rm S} + \sqrt{\eta(R)(G_{\rm S, tr}-1)}\, \hat{c}_{\rm S, tr}^\dagger \notag\\
	&+ \sqrt{1 - \eta(R)}\, \hat{a}_{\rm env}. \label{Eq. annihilation_incident field}
\end{align}
{\color{black}From Eqs.\eqref{Eq. Received number} and \eqref{Eq. annihilation_incident field}, one can obtain} the total number of photons incident on the radar receiver—including contributions from both the entangled signal and {\color{black}the} added noise {\color{black}as follows}
\begin{equation}
	N_{\rm inc} = \eta(R) G_{\rm S, tr} N_s + N_{\rm n, inc}, \label{Eq. Received Photons}
\end{equation}
{\color{black}where}
\begin{equation}
	N_{\rm n, inc} = [1 - \eta(R)] N_{\rm env} + \eta(R)(G_{\rm S, tr} - 1)(N_{\rm S, tr}^{\rm amp} + 1), \label{Eq. noise_incident}
\end{equation}
represents the total added noise into the incident field. 
{\color{black}Thus, the} corresponding power of the incident field can be calculated as
\begin{align}
	P_{\rm inc} &= h f_{\rm S} N_{\rm inc} B_{\rm a} \notag\\
	&= P_{\rm S, inc} + P_{\rm n, inc}, \label{Eq. power_incident}
\end{align}
{\color{black}where} $P_{\rm S, inc}$ is the power of the {\color{black}back-scattered} entangled signal reaching the receiver antenna, {\color{black}and is given by}
\begin{equation}
	P_{\rm S, inc} = \eta(R) G_{\rm S, tr} h f_{\rm S} N_s B_{\rm a}.
\end{equation}
Moreover, {\color{black}the noise power at the antenna $P_{\rm n, inc}$ is} given by
\begin{equation}
	P_{\rm n, inc} = h f_{\rm S} N_{\rm n, inc} B_{\rm a}.
\end{equation}
It is crucial to emphasize that the radar receiver must be sensitive to the entangled portion of the incident field, {\color{black}i.e., $P_{\rm S, inc}$}. If the power of this component falls below the PDL, no photocurrent will be generated, and consequently, no information about the target can be extracted.

{\color{black}In Fig.~\ref{Fig9b}, we have demonstrated the total incident power, $P_{\rm inc}$, alongside the power of the entangled signal, $P_{\rm S, inc}$, that reached the receiver. In this figure, the PDL of various microwave detection scenarios is illustrated. }  
The system parameters used in {\color{black}Fig.~\ref{Fig9b} are the same as} summarized in Table~\ref{tab1}, {\color{black}with the target RCS as} $\sigma = 0.5~\mathrm{m}^2$. 
{\color{black}Here,} the red, yellow, blue, and green shaded {\color{black}regions} correspond to the operational detection {\color{black}regimes} for scenarios that utilize {\color{black}the} quantum transducer combined with {\color{black}the} SOPD, SMPD, {\color{black}the} low-temperature and {\color{black}the} {\color{black}room-temperature} classical antenna, respectively. Incident power below the red-shaded region {\color{black}is} undetectable with {\color{black}the} current detection technologies.

{\color{black}According to Fig.~\ref{Fig9b}, for the case where $N_s = 0.05$, by employing the room-temperature antenna, the considered system can detect the target with the maximum range of $4.8~\rm m$.}
However, by employing low-temperature classical antennas operating at approximately $30~\mathrm{K}$, the detection range extends to roughly $8.45~\mathrm{m}$.
Utilizing the state-of-the-art SMPD, such as the one reported in {\color{black}Ref.~ \cite{pallegoix2025enhancing}, the detection range increases up to $1.19~\mathrm{km}$,} {\color{black}an increase of more than 200 times relative to the room-temperature antenna case}.
{\color{black}In addition, the detection scenario based on the quantum transducer introduced in Ref.~\cite{zhao2025quantum} enables the detection of weak received signals from the target at ranges up to $2.15~\mathrm{km}$.} 
Here, to obtain these results we have assumed $N_s = 0.5$.

It is important to note that {\color{black}the maximum detection range obtained in} Eq.~\eqref{rangeEqNR_final_2} assumes an ideal microwave receiver, and {\color{black}did not consider the limitation imposed by the PDL.}
{\color{black}In order to} account this, {\color{black}in addition to the result obtained in Eq.~\eqref{rangeEqNR_final_2}, we should consider} the range at which the power of the entangled signal at the receiver becomes equivalent to the PDL of the receiver, denoted {\color{black}by} $R_{\rm PDL}$. {\color{black}One can obtain} this range by solving the equation $P_{\rm S, inc} = P_{\rm PDL}$ for $R_{\rm PDL}$, {\color{black}which yields}
\begin{equation}
	R_{\rm PDL} = \frac{20}{\ln(10)\,\gamma} \, W_0 \left[ \frac{\ln(10)\,\gamma}{20} \left( \frac{\sigma G A_e G_{\rm S, tr} h f_{\rm S} N_s B_a}{ (4\pi)^2 P_{\rm PDL}} \right)^{1/4} \right]. \label{Eq. R_PDL}
\end{equation}
{\color{black}As a consequence, for {\color{black}a} quantum-entangled noise radar}, the experimentally achievable detection range is given by
\begin{equation}
	R_{\rm max} \equiv \min \left\{ R_{\rm max}^{\rm QI},\, R_{\rm PDL} \right\}. \label{Eq. extreme range}
\end{equation}
{\color{black}Note that to completely exploit the advantages of quantum-entangled noise radars predicted by Eq.~\eqref{rangeEqNR_final_2}, the microwave detection technology must satisfy the condition $R_{\rm PDL} \ge R_{\rm max}^{\rm QI}$. Therefore, an efficient microwave detection scenario for a quantum-entangled noise radar is that for which this condition is satisfied. Otherwise, the detection range is limited to $R_{\rm PDL}$, which results in a reduction of the REF compared to its classical counterpart. However, it may still be capable of detecting targets at kilometer-range distances.}

\subsubsection{Application-Specific Suitability} 
\begin{figure}[!htb]
	\centering
	\includegraphics[width=250pt]{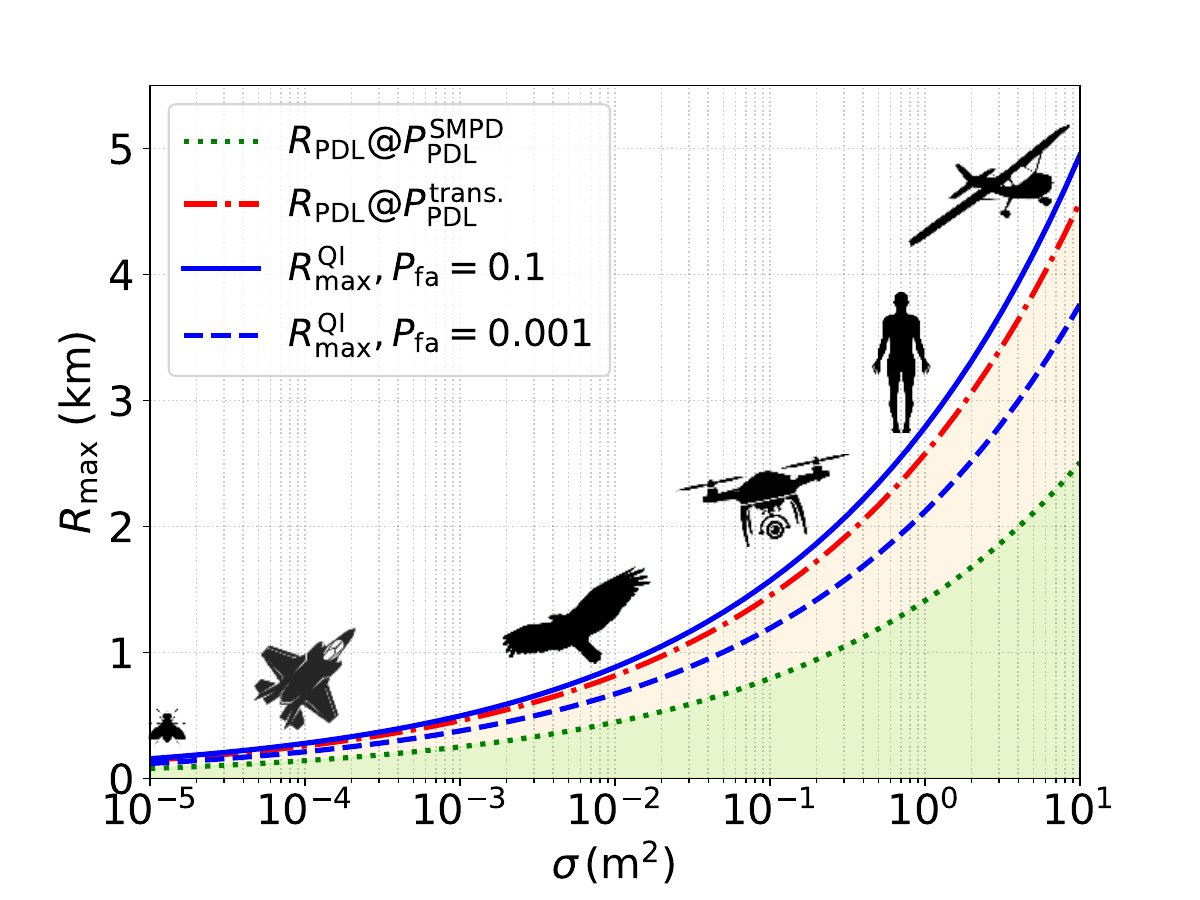}
	\caption{(Color online) {\color{black}The} maximum detection range of {\color{black}the} quantum-entangled noise radar, $R_\mathrm{max}^\mathrm{QI}$, in terms of the object's RCS, $\sigma$, for $N_s=0.05$. {\color{black}The other system parameters} are {\color{black}the same as} in Table.~\ref{tab1}. 
	}
	\label{Fig10}
\end{figure}

Experimental implementations of quantum-entangled noise radars have been restricted to {\color{black}operation} {\color{black}inside controlled laboratory environments. Up to now, no practical systems have been realized} for specific real-world applications {\color{black}yet}. To provide a perspective on potential deployments {\color{black}in the future}, we have plotted the maximum detection range of the quantum-entangled noise radar in terms of the target RCS in Fig.~\ref{Fig10}. The system parameters are {\color{black}the same as} listed in Table~\ref{tab1} {\color{black}with} $N_s = 0.05$. 
{\color{black}Here, the} schematic icons representing different targets (i.e., insect, stealth aircraft, bird, commercial drone, human body, and light aircraft) {\color{black}are included to provide a quick visual sense for readers}.
{\color{black}Moreover, the parameter $R_{\rm PDL}$ for the microwave detection technologies based on the quantum transducer and SMPD {\color{black}are} illustrated by the {\color{black}red} dot-dashed and {\color{black}green} dotted curves, respectively.}

%
As shown in Fig.~\ref{Fig10}, {\color{black}in the} search mode {\color{black}with} $P_{\rm fa} = 0.1$, the maximum detection range predicted by Eq.~\eqref{rangeEqNR_final_2} exceeds $R_{\rm PDL}$ when employing the detection scenario based on {\color{black}the} quantum transducer. Consequently, as indicated by Eq.~\eqref{Eq. extreme range}, the maximum achievable detection range is constrained by $R_{\rm PDL}$. However, the difference between the two is negligible.
In {\color{black}the} track mode {\color{black}with} $P_{\rm fa} = 0.001$, the maximum detection range predicted by Eq.~\eqref{rangeEqNR_final_2} remains below $R_{\rm PDL}$ for the same detection scenario. Therefore, in {\color{black}the} track mode, Eq.~\eqref{rangeEqNR_final_2} accurately describes the achievable detection range of the {\color{black}considered} quantum-entangled noise radar.
In contrast, for the detection scenario {\color{black}that utilizes} {\color{black}the} SMPD, the parameter $R_{\rm PDL}$ is significantly lower than the range predicted by Eq.~\eqref{rangeEqNR_final_2} in both {\color{black}the} search and track modes. As a result, the ultimate detection range in this case is determined by $R_{\rm PDL}$. Consequently, even with {\color{black}the} state-of-the-art SMPDs featuring {\color{black}a} {\color{black}DCR as low as $30~\rm Hz$} \cite{pallegoix2025enhancing}, the system performance is significantly restricted. As illustrated in Fig.~\ref{Fig9a}, classical antennas, even when cooled to low temperatures, exhibit $R_{\rm PDL} \simeq 8.45~\rm m$, which is substantially {\color{black}lower than} the maximum detection range predicted by Eq.~\eqref{rangeEqNR_final_2}. This discrepancy effectively nullifies the advantages of quantum-entangled radar systems. This may explain the limited range, of approximately $1~\rm m$, of the quantum-entangled radar system implemented in Ref. \cite{balaji2019receiver}.	 

In summary, to effectively employ the performance advantages offered by quantum-entangled noise radars, the use of a microwave detection system based on the quantum transducer combined with a high-efficiency SOPD is strongly recommended. {\color{black}It should be noted that the following analysis is based on the assumption of using the quantum {\color{black}transducer-based} microwave detection technology}.

Figure~\ref{Fig10} demonstrates that the considered quantum-entangled {\color{black}noise} radar can detect a stealth aircraft with {\color{black}an} RCS around $\sigma \approx 10^{-4}~\mathrm{m}^2$ at ranges up to $271~\rm m$ ($206~\rm m$) in the search (track) mode. A typical stealth aircraft travels at speeds of $200-700~\rm m/s$. {\color{black}By considering the minimum warning time required for its detection as 60 seconds, then the necessary detection range is obtained to be $12-42~\rm km$, which greatly exceeds the quantum-entangled radar’s maximum detection range}. Consequently, the current quantum-entangled {\color{black}noise} radar configuration {\color{black}is} inadequate for early-warning applications for high-speed stealth aircraft.

However, for typical commercial drones with {\color{black}an RCS of} the order of $\sigma \approx 0.5~\rm m^2$ and moderate speeds of approximately $20-30~\rm m/s$, the maximum detection range of the {\color{black}considered} quantum-entangled {\color{black}noise} radar is approximately $2.28~\rm km$ in {\color{black}the} search mode and $1.7~\rm km$ in {\color{black}the} track mode. It is comparable to or exceeds the required warning distance for such targets. Therefore, the quantum-entangled {\color{black}noise} radar under investigation {\color{black}could be} suitable for detecting and tracking common commercial drones {\color{black}in real-world applications}.

For {\color{black}a} light aircraft with typical speeds of $60-70~\rm m/s$ and {\color{black}an} RCS of approximately $\sigma \approx 10~\rm m^2$, the maximum detection range of the quantum-entangled {\color{black}noise} radar is $4.8~\rm km$ in {\color{black}the} search mode and $3.65~\rm km$ in {\color{black}the} track mode. This performance is marginally consistent with the operational range of precision approach radars (PAR) used at airports. Accordingly, quantum-entangled {\color{black}noise} radars may have potential applications in airport surveillance systems.

Regarding {\color{black}the} other targets illustrated in Fig.~\ref{Fig9}, such as birds and humans, the quantum-entangled {\color{black}noise} radar demonstrates acceptable maximum detection ranges. Consequently, potential applications for quantum-entangled noise radars include radar-based non-contact human vital sign detection \cite{zhang2023photonic} as well as studies in aeroecology and aerobiology \cite{gauthreaux2003radar}.

{\color{black}In this study,} we have focused solely on evaluating the operational performance of the quantum-entangled {\color{black}noise} radar based on {\color{black}the} current state-of-the-art technologies. The system performance has been {\color{black}examined} for specific applications to assess their suitability, without considering practical challenges such as implementation complexity, cost, size, weight, and power consumption (SWaP) \cite{defenceReportCanada, balajiSnapshot2020}. These practical issues are beyond the scope of the present {\color{black}work} and warrant comprehensive analysis tailored to each individual application.

\subsubsection{ Entanglement in Quantum-Entangled Noise Radars }
We {\color{black}are now going to} revisit the challenge concerning the entanglement between the {\color{black}back-scattered} signal {\color{black}from the target at a range $R$} and the retained idler. {\color{black}In order to address this issue, one} {\color{black}needs to} evaluate {\color{black}an appropriate entanglement measure} {\color{black}between the signal and idler modes}.
{\color{black}Here,} Simon’s separability criterion is employed for this purpose. As demonstrated in Appendix~\ref{appendix_Simon}, the signal and idler modes generated at the source, represented by Eq.~\eqref{psi_TMSV}, are shown to be inseparable (i.e., entangled) for any photon number {\color{black}per mode} $N_s > 0$.
Nonetheless, {\color{black}it vanishes due} to signal amplification, free-space propagation, and detection at the receiver (cf. Appendix~\ref{appendix_Simon}). This observation raises an important question: if the quantum entanglement is sufficiently {\color{black}fragile and} completely destroyed during {\color{black}the} free-space propagation, what then underlies the advantage of {\color{black}the} quantum-entangled noise radars over the classical ones?

{\color{black}To clarify the apparent paradox, it must be remembered that the absence of entanglement is not equivalent to classicality \cite{Weedbrook2016discord}. 
As} the initial entanglement is entirely eradicated during propagation through the free-space channel, the quantum advantage {\color{black}might still persisted} due to the presence of quantum correlations beyond entanglement. {\color{black}Particularly, it can} be attributed to quantum discord, {\color{black}which is a} more robust quantum correlation than entanglement. {\color{black}Thus, the quantum discord can be seen as a quantum resource of quantum correlation} in such scenarios \cite{Weedbrook2016discord}.

Quantum discord is defined as the difference between the total correlations within a quantum state, as quantified by the quantum mutual information, and and the classical correlations of that state. It corresponds to the maximum amount of shared randomness accessible via local measurements and one-way classical communication \cite{Devetak2014discord, Pirandola2014discord}. In essence, the residual quantum correlations that remain despite the loss of entanglement are responsible for the improved performance of entanglement{\color{black}-based} target detection schemes {\color{black}versus the classical target detection}.


\section{Summary, conclusion remarks and outlooks} \label{sec6.Results and Discussion}

In this paper, we {\color{black}have presented} an analysis of the operational performance of {\color{black}the} quantum single-photon direct-detection radars (see Sec.~\ref{sec2:direct_detection_radars}) and quantum-entangled noise radars (see Sec.~\ref{sec3.NoiseRdar}), focusing on their maximum detection range. 
For both radar types, we {\color{black}have calculated the} explicit expressions for the maximum detection range in terms of the Lambert W function. 
Our analysis {\color{black}has provided} a comprehensive model that incorporates all {\color{black}the} relevant system and environmental parameters.

Analogous to single-photon LiDAR systems, we {\color{black}have explored} the feasibility of employing single microwave-photon detectors as the radar receiver to realize single-photon radar systems. Based on recent advances in SMPD technology, we {\color{black}have demonstrated} that implementing a single-photon radar with {\color{black}the} transmit power {\color{black}of} the order of a few milli-watts and the detection range in the order of a few kilometers is feasible (see Sec.\ref{sec2:direct_detection_radars}). This development holds promise for the realization of low-power, high-performance {\color{black}single-photon} radar systems for practical field applications. Nevertheless, further advancement and optimization of SMPD technology {\color{black}is} necessary to achieve this goal.

In {\color{black}the case of} quantum-entangled noise radars, we have derived an explicit expression for the maximum detection range in terms of the Lambert W function. {\color{black}This allows us to define an} effective threshold SNR for quantum-entangled noise radars (see Eq.~\eqref{SNR_th_NR}). {\color{black}This} suggests that {\color{black}a noise radar} can be interpreted {\color{black}effectively as a} direct-detection radar with a reduced threshold SNR.

To demonstrate the performance advantage of quantum-entangled noise radars {\color{black}with respect to} their classical counterparts, we {\color{black}have also introduced the $\mathrm{REF}$ parameter}. {\color{black}We have also derived a rule-of-thumb approximation for the $\rm REF$ in terms of} the quantum advantage $Q_{\rm adv}$, which depends solely on the mean photon number {\color{black}per mode}, $N_s$ (see Eq.~\eqref{REF_2}). Moreover, we have linked REF to SNR enhancement from {\color{black}the} engineering point of view.  
{\color{black}In addition, we have classified the} quantum-entangled noise radars into two operational modes {\color{black}based on the} {\color{black}false-alarm probability: the search and track modes}.

{\color{black}In this paper, a} quantitative evaluation of the maximum detection range is provided using feasible system parameters {\color{black}based on the current} state-of-the-art technologies (see Table~\ref{tab1}).
{\color{black}Our} results show that {\color{black}the quantum-entangled noise radars} have the potential to detect small {\color{black}targets} with {\color{black}the RCS $\sigma = 0.1~\mathrm{m}^2$ up to a range of} $2~\mathrm{km}$ in {\color{black}the} track mode with $P_{\rm fa} = 0.001$. {\color{black}More importantly, we have shown that the target detection occurs} within a reasonable integration time, and {\color{black}the system shows a} substantial range enhancement factor of $\mathrm{REF} = 2.14$ (see Fig.~\ref{Fig8}).
We {\color{black}have} also compared the chosen system parameters with those reported in some of the most notable experimental implementations of {\color{black}the} quantum-entangled noise radars to date (see Table~\ref{tab3}). This comparison demonstrates that our considered {\color{black}system} parameters are achievable {\color{black}via} current technologies.

Additionally, we have addressed several practical challenges related to the operation and application of {\color{black}the} quantum-entangled noise radars (see Subsec.~\ref{subsec3.5-challenges}). In particular, we have explored the potential use of these radars for detecting various targets, such as stealth aircraft, birds, commercial drones, human bodies, and light aircraft. For stealth aircraft, by comparing the radar's maximum detection range with the minimum warning range required for such high-speed {\color{black}and} low-RCS targets, {\color{black}one can} conclude that quantum-entangled noise {\color{black}radars} {\color{black}under} current technological constraints {\color{black}are not yet} suitable for this application. {\color{black}However, they have} strong potential for detecting commercial drones at urban distances. Furthermore, {\color{black}they} might be used as precision approach radars in airports (see Fig.~\ref{Fig9}).

{\color{black}By} considering the inherently low power of the transmitted entangled signal in quantum-entangled noise radars, we {\color{black}have also examined} whether the power of the back{\color{black}-scattered} signal is sufficient for detection {\color{black}with} current radar receivers. To this end, we {\color{black}have} compared the received signal power with the physical detection limit of modern microwave {\color{black}receiver} antennas and state-of-the-art SMPDs. Our analysis shows that, for instance, the maximum range at which the received signal power {\color{black}reaches the PDL} is approximately $2.23~\mathrm{km}$ for $N_s = 0.5$. To overcome this limitation, one can use high-efficiency quantum microwave-to-optical transducers, such as those based on quantum electro-optomechanical systems \cite{vitaliCoverter2023}, to coherently convert the back{\color{black}-scattered} microwave signals into optical ones. This approach {\color{black}employs highly efficient SOPDs for detecting optical signals}.

From {\color{black}an} engineering and practical standpoint, several key directions can be pursued {\color{black}to achieve long-range quantum-entangled noise radars as follows}:
\begin{itemize}
	\item \textit{Broadband entangled microwave sources}: Increasing the operational bandwidth of entangled microwave sources {\color{black}is essential} for enhancing both the {\color{black}maximum detection} range and quantum advantage. This can be achieved through platforms based on electro-optomechanical (EOM) systems, superconducting circuits, or spin-based systems such as nitrogen-vacancy centers.
	
	\item \textit{Photonics-based radar platforms}: Integrating quantum-entangled noise radar architectures with photonic radar systems can facilitate more efficient signal detection and post-processing~\cite{ghelfiPhotonicRadar1}. High-efficiency quantum \textit{microwave-to-optics transducers}—such as EOM-based transducers\cite{vitaliCoverter2023}— and {\color{black}efficient RF antennas designed specifically} for the emission of entangled microwave signals in open-air \cite{gonzalez2022coplanar} are critical for enabling this integration.
	
	\item \textit{Room-temperature entanglement sources}: Generating entangled microwave signals at ambient or {\color{black}high} temperatures {\color{black}is possible through utilizing} advanced EOM platforms~\cite{marinCooing2023Room,kippenbergRoomTemperaure2024} or superconducting systems~\cite{microwave_Photon_Emission_Superconducting_Circuits} via quantum control techniques. Exploring bulk or integrated nonlinear optical crystals \cite{aliBBO, aliPPKTP, aliDNA}, NV centers \cite{aliNVradar}, or other room-temperature quantum sources could eliminate the need {\color{black}for} cryogenic systems. Utilizing high-temperature microwave photon-pair sources can significantly reduce system SWaP and cost.
	
	\item \textit{\color{black}Single microwave-photon detectors}: The development of low DCR, high-sensitivity single microwave photon detectors—especially those that {\color{black}can operate} at higher temperatures \cite{rfSensingNV2023, NVreceiverfT2024, NVreceiverPT2024}—would substantially reduce the effective threshold SNR {\color{black}which enhances} {\color{black}the} radar performance.

	\item \textit{Investigate quantum-entangled noise radars based on non-Gaussian entangled states}: The analysis of the maximum detection range presented in this work {\color{black}is restricted} to the use of quantum two-mode Gaussian states {\color{black}to generate} the correlated signal and idler fields~\cite{xu2022metrological, zhang2024quantum}. {\color{black}It seems that} employing non-Gaussian quantum entangled {\color{black}states may} enhance the quantum advantage {\color{black}and also further increase} the maximum detection range.
	
\end{itemize}

\section*{Author Contribution}
AMF as the corresponding author and director of the quantum sensing $ \& $ metrology Lab/group at ICQT has defined and led this project.  
All calculations have been done by HA and rechecked and interpreted by AMF. All numerical calculations and graphs have been plotted by HA. The subjects of the quantum antenna or single-photon microwave detectors for improving the threshold-SNR have been introduced and developed by AMF. The effective-SNR has been defined and developed by HA and has been interpreted by AMF. {\color{black}AMF introduced and developed the engineering parameters and approaches such as REF and so on, and then, HA checked and re-formulated them}. Experimental parameters have been determined and extracted by both authors. Both authors have contributed to the discussion and interpretation of the results as well as writing, editing and revising the manuscript. The idea of investigating the received photon number and Simon criterion are put forwarded by AMF, calculated by HA and then re-checked by AMF. Both authors contributed to answer {\color{black}to the reviewers, and prepare the response letter.}

\section*{Acknowledgments}
The authors thank the Iranian Center for Quantum Technologies (ICQT). {\color{black}AMF would thank the University of Tehran.}

\section*{Author Declarations}
\subsection*{Conflict of Interest}
The authors have no conflicts to disclose.

\section*{Data Availability Statement}
This study is entirely theoretical in nature. No experimental data were generated or analyzed, and thus no data are available for public access.

\appendix

\section{Squeezing origination of microwave QTMS state in JPA \label{appendix_squeezing} } 

It is well-known that in the optical domain, the entangled Gaussian states can be generated by {\color{black}the} SPDC, a 3-wave mixing process in NLCs in which a pump photon {\color{black}with the higher energy} is converted through {\color{black}the} second-order nonlinear interaction with quantum vacuum fluctuations into a pair of signal-idler photons {\color{black}with the lower energy} respecting {\color{black}the} energy-momentum conservation law. Since signal-idler biphoton is created from the same {\color{black}quantum} vacuum, they both share entanglement between each other stronger than any classical correlation.
On the other hand, in the microwave band, phase-preserving JPAs, as a quantum-limited amplifiers that operate in 3-10GHz band with signal-idler frequency which can be adjusted in GHz-band, are responsible for the microwave entanglement generation via the well-known 4-wave mixing process \cite{wilsonProgress2020}. 
In a JPA, two input vacuum signal ports ($ \hat \nu, \hat \mu $) are mixed together and two amplified sideband signal and idler with power gain $ G_{\mathrm{JPA}} $ are generated. The generation of entanglement from uncorrelated vacuum noise inputs in a JPA can be modeled as \cite{wilsonProgress2020} 
\begin{eqnarray}
	&& \hat a_{\mathrm{s}}=\cosh (r) \hat \nu + \sinh (r) \hat \mu^\dagger , \\
	&&  \hat a_{\mathrm{i}}=\sinh (r) \hat \nu^\dagger + \cosh (r) \hat \mu,
\end{eqnarray}
where $ r $ is referred to the squeezing parameter and is related to the amplification gain power as $ G_{\mathrm{JPA}}=\cosh^2(r) $. 
It is simple to show that the signal-idler quadratures of a QTMS state is a zero-mean Gaussian thermal state with average photon number as the so-called photon {\color{black}per} mode $ N_s=[\cosh(2r)-1]/2 $ which immediately implies that signal-idler correlation in the Covariance matrix becomes $ \mathcal{C}_Q=\sinh(2r)/2 $ {\color{black}which} can be greater than its classical counterpart.
Surprisingly, this quantum advantage is responsible for the noise reduction below {\color{black}the} quantum vacuum ($ 1/2 $) in generalized quadratures $ \hat I_-=(\hat I_{\mathrm{s}}-\hat I_{\mathrm{i}})/\sqrt{2} $ and $ \hat Q_+=(\hat Q_{\mathrm{s}}+\hat Q_{\mathrm{i}})/\sqrt{2} $. This leads to quadrature squeezing as {\color{black}follows}
\begin{eqnarray}
	&& {\rm Var}[\hat I_-]=e^{-2r}/2, 
\end{eqnarray}
which has been experimentally realized up to around -12dB \cite{microwaveSquuezing2014}.

\section{Simon's Separability  {\color{black}Criterion} \label{appendix_Simon}}
Here, in this appendix we are going to calculate the Simon's separability {\color{black}criterion} for the signal and idler modes before the transmission of the signal and also after returning  {\color{black}the back-scattered signal from the target} to the receiver in general situation by considering attention, amplification and noise. To this, we starts with the covariance matrix for {\color{black}the} signal and idler modes at the transmitter which is given by \cite{lanzagorta2015low}
\begin{align}
	{\rm Cov}_{\rm tr}=\frac{1}{4}\left( {\begin{array}{*{20}{c}}
			S_1 & 0 & C_q & 0 \\
			0 & S_1 & 0  & -C_q\\
			C_q & 0 & S_2 & 0 \\
			0 & -C_q & 0 & S_2
	\end{array}} \right),\label{eq. A2-1}
\end{align} 
{\color{black}where} $S_1=S_2= 2N_s + 1$, and $C_q = \sqrt{N_s(N_s + 1)}$. By utilizing the annihilation operator of the detected idler and signal fields given in Eqs.~\eqref{a_hat_S_det} and \eqref{a_hat_I}, the covariance matrix for the detected signal and idler fields is obtained as
\begin{align}
	{\rm Cov}_{\rm rec}=\frac{1}{4}\left( {\begin{array}{*{20}{c}}
			S_1' & 0 & C_q' & 0 \\
			0 & S_1' & 0  & -C_q'\\
			C_q' & 0 & S_2' & 0 \\
			0 & -C_q' & 0 & S_2'
	\end{array}} \right), \label{eq. A2. 4}
\end{align}
{\color{black}with}
\begin{align}
	S_1'=&2\big[\eta G_{\rm S} N_\mathrm{s} + G_\mathrm{S, rec} N_\mathrm{env} + G_\mathrm{S,rec}(G_\mathrm{I}-1)(N_\mathrm{S, tr}^\mathrm{amp}+1) \notag\\
	&+(G_\mathrm{S,rec}-1)(N_\mathrm{S, rec}^\mathrm{amp}+1)\big]+1, \label{eq. A2. 5}\\
	S_2'=&2\big[G_\mathrm{I}N_\mathrm{s} 
	+(G_\mathrm{I}-1)(N_\mathrm{I}^\mathrm{amp}+1)\big]+1 \label{eq. A2. 6} \\
	C_q'=& \sqrt{\eta G_\mathrm{s}^\mathrm{amp}G_\mathrm{i}^\mathrm{amp}}C_q, \label{eq. A2. 7}
\end{align}
{\color{black}The parameters in these relations have previously defined in Sec.~\ref{sec3.NoiseRdar}}.
It is worth noting that when there is no amplification, i.e., {\color{black}$G_\mathrm{S, rec}=G_\mathrm{I}=1$}, the covariance matrix given in Eq.~\eqref{eq. A2. 4} reduces to that reported in references \cite{tan2008quantum, lanzagorta2015low} for QTMS radar with no amplification. 

As mentioned in {\color{black}Ref.}~\cite{lanzagorta2015low}, the entanglement between the signal and idler mode can be evaluated through the so-called Simon's separability {\color{black}criterion} as
\begin{align}
	f \ge 0,
\end{align}
in which $f$ is defined as Simon parameter, and for the signal and idler modes represented by the covariance matrix of the form \eqref{eq. A2-1} or \eqref{eq. A2. 4} is defined as
\begin{align}
	f \equiv (S_1 S_2-C_q^2)^2 - (S_1^2 + S_2^2 + 2C_q^2) + 1. \label{eq. A2. 8}
\end{align}
Using this equation, it is straightforward to show that $f$ at the source is given by
\begin{align}
	f_\mathrm{source}=-16N_s(N_s+1), \label{eq. A2. 9}
\end{align}
which is negative for any $N_\mathrm{s}>0$. Therefore, it is concluded that the generated signal and idler modes are inseparable or entangled. However, using Eq.~\eqref{eq. A2. 5}-\eqref{eq. A2. 7} one can obtain $f$ for the detected signal and idler modes as
\begin{align}
	f_\mathrm{det} = (S_1'S_2' - C_q'^2)^2 - (S_1'^2 + S_2' + 2C_q') + 1. \label{eq. A2. 10}
\end{align}
By considering {\color{black}$G_\mathrm{I}=77~\mathrm{dB}$, $G_\mathrm{S,rec}=17~\mathrm{dB}$}, and the {\color{black}other system} parameters {\color{black}as the same as reported} in Table.~\eqref{tab1}, {\color{black}one finds} that $f_\mathrm{det}\gg 1$ for any $R>0$. Therefore, the detected signal and idler modes are not entangled, and {\color{black}what is help us to exploit the entanglement is quantum discord as the quantum resource of residual quantum correlation}.
		


	\end{document}